\documentclass[a5paper,12pt,twoside]{amsbook}
\usepackage{epsfig}
\usepackage{amsmath}
\usepackage{amsfonts}
\usepackage{amssymb}
\usepackage{euscript}
\usepackage{graphicx}
\usepackage{psfrag}
\usepackage{rotating}

\begin{document}

\sloppy



$\qquad$

\vspace{20mm}
{\Large\underline{Annotation}}\\

Currently there is a common belief that the explanation of superconductivity phenomenon lies in understanding the mechanism of the formation of electron pairs.
Paired electrons, however, cannot form a superconducting condensate spontaneously. These paired electrons perform disorderly zero-point oscillations and there are no force of attraction in their ensemble. In order to create a unified ensemble of particles, the pairs must order their zero-point  fluctuations so that an attraction between the particles appears. As a result of this ordering of zero-point oscillations in the electron gas, superconductivity arises.  This model of condensation of zero-point oscillations creates the possibility of being able to obtain estimates for the critical parameters of elementary superconductors, which are in satisfactory agreement with the measured data.  On the another hand, the phenomenon of superfluidity in He-4 and He-3 can be similarly explained, due to the ordering of zero-point fluctuations. It is therefore established that both related phenomena are based on the same physical mechanism.\\



\newpage


\thispagestyle{empty}


\begin{center}
{
\Large

\itshape{ Boris V.Vasiliev}}
\end{center}
\vspace{30mm}

\sloppy

\begin{center}
{{\Huge \bfseries SUPERCONDUCTIVITY\\
\vspace{20mm}
and\\
\vspace{20mm}
SUPERFLUIDITY
}}
\end{center}

\begin{center}
\part{The development of the science of superconductivity and superfluidity}
\end{center}

\chapter{Introduction}

\section{Superconductivity and public}
Superconductivity is a beautiful and unique natural phenomenon that was discovered in the early 20th century.   Its unique nature comes from the fact that superconductivity is the result of quantum laws that act on a macroscopic ensemble of particles as a whole.
 The concept of superconductivity is attractive not only  for circles of scholars, professionals and people interested in physics, but wide educated community.

    Extraordinary public interest in this phenomenon was expressed to the scientific community just after the discovery of high temperature superconductors in 1986. Crowds of people in many countries gathered to listen to the news from scientific laboratories.
 This was  the unique event at this time, when the scientific issue was the cause of such interest not only in narrow circle of professionals but also in the wide  scientific community.

    This interest was then followed by public recognition.  One sign of this recognition is through the many awards of Nobel Prize in physics. This is one area of physical science, where plethoras of Nobel Prizes were awarded. The chronology of these awards follows:\\

1913: Heike Kamerlingh-Onnes was awarded the Nobel Prize in Physics for the discovery of superconductivity;\\

1962: Lev Landau was awarded the Nobel Prize in Physics in ”for the pioneering theories for condensed matter, especially liquid helium,” or in other words, for the explanation of the phenomenon of superfluidity.\\

1972: John Bardeen, Leon N.  Cooper and J. Robert Schrieffer shared the Nobel Prize in Physics “for the development of the theory of superconductivity, usually called the BCS theory.”\\

1973: Brian D. Josephson was awarded the Nobel Prize in Physics in  ”for the theoretical predictions of the properties of the superconducting current flowing through the tunnel barrier, in particular, the phenomena commonly known today under the name of the Josephson effects”\\

1978: Pyotr Kapitsa was awarded the Nobel Prize in Physics, ”for his basic inventions and discoveries in the area of low-temperature physics,” that is, for the discovery of superfluidity.\\

1987: Georg Bednorz and Alex Muller received the Nobel Prize in Physics for “an important breakthrough in the discovery of superconductivity in ceramic materials.”\\

2003: Alexei Abrikosov, Vitaly  Ginzburg and Anthony Legett received the Nobel Prize in Physics for “pioneering contributions to the theory of superconductors and superfluids”.\\

Of course, the general attention to superconductivity is caused not just by its unique scientific beauty, but in the hopes for the promise of huge technological advances. These technological advances pave the way, creating improved technological conditions for a wide range of applications of superconductivity in practical societal uses:
maglev trains, lossless transmission lines, new accelarators, devices for medical diagnostics and devices based on highly sensitive SQUIDs.\\

Because of these discoveries, it may now look like there is no need for the development of the fundamental theory of superconductivity at all.  It would seem that the most important discoveries in superconductivity have been already made, though more or less randomly.

Isaac Kikoine, a leading Soviet physicist, made a significant contribution to the study of superconductivity on its early on stage.\footnote{I. Kikoine`s study of the gyromagnetic effect in the superconductor in the early 1930s led him to determination of the gyromagnetic factor of the superconducting carriers.}

He used to say, whilst referring to superconductivity, that many great scientific discoveries was made by Columbus method. This was when, figuratively speaking, “America” was discovered by a researcher who was going to “India”. This was a way by which Kamerlingh-Onnes came to his discovery of superconductivity, as well as a number of other researchers in this field.

Our current understanding of superconductivity suggests that it is a specific physical discipline.
It is the only area of physics where important physical quantities equals exactly to zero. In other areas of physics small and very small values exist, but there are none which are exactly zero.  A property can be attributed to the ‘zero’ value, in the sense there is a complete absence of the considered object. For example, one can speak about a zero neutrino mass, the zero electric charge of neutrons, etc., but these terms have a different meaning.\\

Is the electrical resistance of superconductors equal to zero?\\

To test this, H. Kamerlingh-Onnes froze a circulating current in a hollow superconducting cylinder. If the resistance was still there, the  magnetic field this current generated would reduced. It was almost a hundred years ago when Kamerlingh Onnes even took this cylinder with the frozen current from Leiden to Cambridge to showcase his findings. No reduction of the field was found.

Now it is clear that resistance of a superconductor should be exactly equal to zero.  This follows the fact that the current flow through the superconductor is based on a quantum effect. The behavior of electrons in a superconductor are therefore governed by the same laws as in an atom.  Therefore, in this sense, the circulating current over a superconductor ring is analogous to the movement of electrons over their atomic orbits.

Now we know about superconductivity, more specifically — it is a quantum phenomenon in a macroscopic manifestation.\\

It seems that the main obstacles in the way of superconductivity`s applications in practice are not the theoretical problems of its in-depth study, but more a purely technological problem. In short, working with liquid helium is too time-consuming and costly and also the technology of nitrogen level superconductors has not yet mastered.

The main problem still lies in correct understanding the physics of superconductivity. Of course, R. Kirchhoff was correct saying that there is nothing more practical than good theory.  Therefore, despite the obvious and critical importance of the issues related to the application of superconductors and challenges faced by their technology, they will not be considered.

The most important task of the fundamental theory of superconductivity is to understand the physical mechanisms forming the superconducting  state. That is, to find out how the basic parameters of superconductors depend on other physical properties.  We also need to overcome a fact that the current theory of superconductivity  could not explain — why  some superconductors  have been observed at critical temperature in a critical field.

These were the facts and concepts that have defined our approach to this consideration.

 In the first part of it, we focus on the steps made to understand the phenomenon of
superconductivity and the problems science has encountered along the way.

 The second part of our consideration focuses on the explanation of superconductivity, which has been described as the consequence of ordering of zero-point  fluctuations of electrons
and that are in satisfactory agreement with the measured data.

The phenomenon of superfluidity in He-4 and He-3 can be similarly explained, due to ordering of zero-point fluctuations.

Thus, it is important that both related phenomena are based on the same physical mechanism.
\newpage

\section{Discovery of superconductivity}
\begin{figure}
\hspace{1.5cm}
\includegraphics[scale=1]{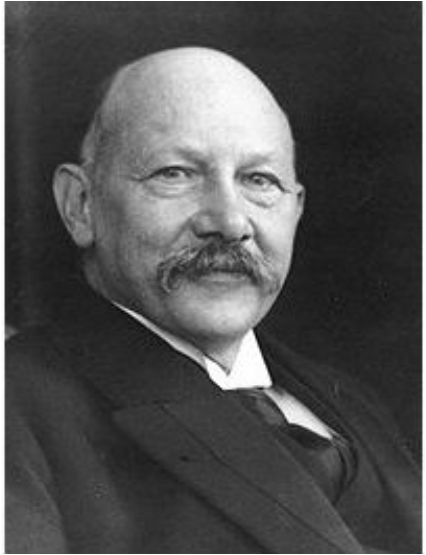}
\caption{Heike Kamerling-Onnes}
\end{figure}

At the beginning of the twentieth century, a new form of scientific research of appeared in the world.  Heike Kamerling-Onnes was one of first  scientists  who used the industry for the service of physics.
His research laboratory was  based on the present  plant of freeze consisting of refrigerators which  he developed.
This industrial approach gave him complete benefits of the world ”monopoly” in studies at low temperatures for a long time (15 years!).
Above all, he was able to carry out his solid-state studies at liquid helium (which boils at atmospheric pressure at 4.18 K). He was the first who creates liquid helium in 1908,then he began his systematic studies of the electrical resistance of metals.
It was known from earlier experiments that the electrical resistance of metals decreases with decreasing temperature. Moreover, their residual resistance turned out to be smaller if the metal was cleaner. So the idea arose to measure this dependence in pure platinum and gold. But at that time, it was impossible to get these metals sufficiently clean. In those days, only mercury could be obtained at a very high degree of purification by method of repeated distillation. The researchers were lucky. The superconducting transition in mercury occurs at 4.15K, i.e. slightly below the boiling point of helium. This has created sufficient conditions for the discovery of superconductivity in the first experiment.

One hundred years ago, at the end of November 1911, Heike Kamerlingh Onnes submitted the article \cite{Onnes} where remarkable phenomenon of the complete disappearance of electrical resistance of mercury, which he called superconductivity, was described.
Shortly thereafter, thanks to the evacuation of vapor, H. Kamerling-Onnes and his colleagues discovered superconductivity  in tin and then in other metals, that were not necessarily very pure. It was therefore shown that the degree of cleanliness has little effect on the superconducting transition.

Their discovery concerned the influence of magnetic fields on superconductors.  They therefore determined the existence of the two main criteria of superconductors: the critical temperature and the critical magnetic field.\footnote{Nobel Laureate V.L.Ginzburg gives in his memoirs the excellent description of events related to the discovery of superconductivity.
He drew special attention to the ethical dimension associated with this discovery.
 Ginsburg wrote \cite{Gins}:
 "The measurement of the mercury resistance was held Gilles Holst.
He was first who observed superconductivity in an explicit form.
He was the qualified physicist (later he was the first director of Philips Research Laboratories and professor of Leiden University).
But his name in the Kamerling-Onnes's publication  is not even mentioned.
As indicated in \cite{deNobel}, G.Holst itself, apparently, did not consider such an attitude Kamerling-Onnes unfair and unusual.
The situation is not clear to me, for our time that is very unusual, perhaps 90 years ago in the scientific community mores were very different."}

The physical research at low temperature was started by H. Kamerling-Onnes and has now been developed in many laboratories around the world.

 But even a hundred  years later, the general style of work in the Leiden cryogenic laboratory created by H. Kamerling-Onnes, including the reasonableness of its scientific policy and the power of technical equipment, still impress specialists.

\chapter[Basic milestones]{Basic milestones in the study of superconductivity}\label{Ch2}
The first twenty two years after the discovery of superconductivity, only the Leiden laboratory of H. Kamerling-Onnes engaged in its research.  Later  helium liquefiers began to appear in other places, and other laboratories were began to study superconductivity.
The significant milestone on this way was the discovery of absolute diamagnetism effect of superconductors.
Until that time, superconductors were considered as ideal conductors. W. Meissner and R. Ochsenfeld \cite{Meis} showed in 1933, that if a superconductor is cooled below the critical  temperature in a constant and not very strong magnetic field, then this field is pushed out from the bulk of superconductor.
The field is forced out by undamped currents that flow across the surface.\footnote{Interestingly, Kamerling-Onnes was searching for this effect and carried out the similar experiment almost twenty years earlier.
The liquefaction of helium was very difficult at that time so he was forced to save on it and  used a thin-walled hollow ball of lead  in his measurements. It  is easy to "freeze" the magnetic field in thin-walled sphere and with that the Meissner effect would be masked.}

\section{The London theory}\label{London-theory}
\subsection{}
\begin{figure}
\hspace{1.5cm}
\includegraphics[scale=0.7]{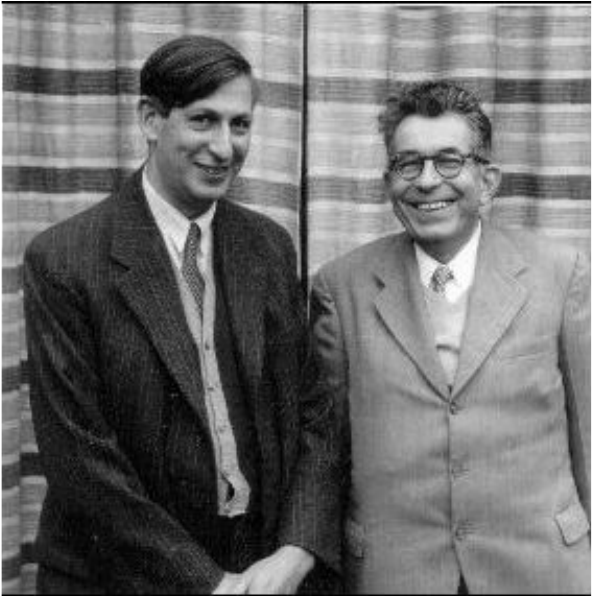}
\caption {Brothers Heinz and Fritz London}
\end{figure}
The great contribution to the development of the science of superconductors was made by brothers Fritz and Heinz London. They offered its first phenomenological theory.
Before the discovery of the absolute diamagnetism of superconductors, it was thought that superconductors are absolute conductors, or in other words, just metals with zero resistance. At a first glance, there is no particulary difference in these definitions.
If we consider a perfect conductor in a magnetic field, the current will be induced onto its surface and will extrude  the field, i.e. diamagnetism will manifest itself.
But if at first we magnetize the sample by placing it in the field, then it will be cooled, diamagnetism should not occur.
However, in accordance with the Meissner-Ochsenfeld effect, the result should not depend on the sequence of the vents.  Inside superconductors the resistance is always:
\begin{equation}
\rho=0,
\end{equation}
and the magnetic induction:
 \begin{equation}
B=0.\label{B0}
\end{equation}

In fact, the London theory \cite{London} is the attempt to impose these conditions on Maxwell’s equations.

The consideration of the London penetration depth is commonly accepted
 (see for example \cite{Ketterson}) in several steps:

\underline{Step 1}.
Firstly, the action of an external electric field on free electrons is considered. In accordance with Newton`s law, free electrons gain acceleration in an electric field $\mathbf{E}$:
\begin{equation}
\mathbf{a}=\frac{e{\mathbf{E}}}{m_e}.\label{aE}
\end{equation}
The directional movement of the "superconducting" \ electron gas with the density
  $n_s$  creates the current with the density:
\begin{equation}
\mathbf{j}=en_s\mathbf{v},
\end{equation}
where $\mathbf{v}$ is the carriers velocity.
After differentiating the time and substituting this in Eq.(\ref{aE}),  we obtain the first London equation:
\begin{equation}
\frac{d}{dt}\mathbf{j}=en_s\mathbf{a}=\frac{n_s e^2}{m_e}\mathbf{E}.\label{Lo1}
\end{equation}

\underline{Step 2}.
After application of operations rot to both sides of this equation and by using Faraday’s law of electromagnetic induction $rot\mathbf{E}=-\frac{1}{c}\frac{d\mathbf{B}}{dt}$,  we then acquire the relationship between the current density and magnetic field:
\begin{equation}
\frac{d}{dt}\left(rot~\mathbf{j}+\frac{n_s e^2}{m_e c}\mathbf{B}\right)=0.\label{L11}
\end{equation}

\underline{Step 3.}
 By selecting the stationary solution of  Eq.(\ref{L11})
 \begin{equation}
rot~\mathbf{j}+\frac{n_s e^2}{m_e c}\mathbf{B}=0,\label{L12}
\end{equation}
and after some simple transformations, one can conclude that there is a so-called  ‘London penetration depth’ of the magnetic field in a superconductor:
\begin{equation}
\Lambda_L=\sqrt{\frac{m_e c^2}{4\pi e^2 n_s}}.\label{lamb}
\end{equation}

\vspace{0.5cm}

\subsection{The London penetration depth and the density of superconducting carriers}
One of the measurable characteristics of superconductors is the London penetration depth, and for many of these superconductors it usually equals to a few hundred Angstroms \cite{Linton}.  In the Table (\ref{L1T}) the measured values of $\lambda_L$ are given in the second column.

\vspace{0.5cm}
\begin{table}
\centering
\hspace{-0.5cm}
\begin{tabular}{||c|c|c|c|c||} \hline\hline
                 & $\lambda_L$,$10^{-6}$cm &      $n_{s}$       &    $n_e$          & \\
  super-& measured                &    according to     &  in accordance  &$n_s/n_e$\\
        conductors & \cite{Linton}           &     Eq.(\ref{lamb}) & with Eq.(\ref{lF}) &\\\hline
  Tl             &   9.2                  &       $3.3\cdot 10^{21}$  &$1.4\cdot 10^{23}$ & 0.023    \\
  In             &   6.4                   &      $6.9\cdot 10^{21}$  &$3.0\cdot 10^{23}$ & 0.024     \\
  Sn             &   5.1                   &      $1.1\cdot 10^{22}$  &$3.0\cdot 10^{23}$ & 0.037     \\
  Hg             &   4.2                   &      $1.6\cdot 10^{22}$  &$4.5\cdot 10^{22}$ & 0.035 \\
  Pb             &   3.9                   &      $1.9\cdot 10^{22}$  &$1.0\cdot 10^{24}$ & 0.019     \\\hline\hline

\end{tabular}
\caption{The London penetration depth and the density of
superconducting carriers}\label{L1T}
\end{table}


\vspace{0.5cm}

If we are to use this experimental data to calculate the density of superconducting carriers $n_s$ in accordance with the Eq.(\ref{lamb}),  the results would about two orders of magnitude larger (see the middle column of Tab.(\ref{L1T}).

Only a small fraction of these free electrons can combine into the pairs. This is only applicable to
the electrons that energies lie within the thin strip of the energy spectrum near $\mathcal{E}_F$.  We can therefore expect that the concentration of superconducting carriers among all free electrons of the metal should be at the level $\frac{n_s}{n_e}\approx 10^{-5}$ (see Eq.(\ref{n0-ne})). These concentrations, if calculated from Eq.(\ref{lamb}), are seen to be  about two orders of magnitude higher (see last column of the Table (\ref{L1T})).
Apparently, the reason for this discrepancy is because of the use of a nonequivalent transformation. At the first stage in Eq.(\ref{aE}),  the straight-line acceleration in a static electric field is considered. If this moves, there will be no current circulation. Therefore, the application of the operation rot in Eq.(\ref{L11})  in this case is not correct. It does not lead to  Eq.(\ref{L12}):
\begin{equation}
\frac{rot~\mathbf{j}}{\frac{n_s e^2}{m_e c}\mathbf{B}}=-1,
\end{equation}
but instead, leads to a pair of equations:
\begin{equation}
\begin{array}{l}
rot~\mathbf{j}=0\\
\frac{n_s e^2}{m_e c}\mathbf{B}=0
\end{array}
\label{L13}
\end{equation}
and to the uncertainty:
\begin{equation}
\frac{rot~\mathbf{j}}{\frac{n_s e^2}{m_e c}\mathbf{B}}=\frac{0}{0}.
\end{equation}

The correction of the ratio of the London’s depth with the density of superconducting carriers will be given in section (\ref{LondonE}).
\newpage

\section{The Ginsburg-Landau theory}
The London phenomenological theory of superconductivity does not account for the quantum effects.

The theory proposed by V.L. Ginzburg and L.D. Landau \cite{GL} in the early 1950’s, uses the mathematical formalism of quantum mechanics. Nevertheless, it is a phenomenological theory, since it does not investigate the nature of superconductivity, although it qualitatively and quantitatively describes many aspects of characteristic effects.

To describe the motion of particles in quantum mechanics one uses the wave function $\Psi (r,t)$, which characterizes the position of a particle in space and time. In the GL-theory, such a function is introduced to describe the entire ensemble of particles and is named the parameter of order. Its square determines the concentration of the superconducting particles.

At its core, the GL-theory uses the apparatus, which was developed by Landau, to describe order-disorder transitions (by Landau’s classification, it is transitions of ‘the second kind’).  According to Landau, the transition to a more orderly system should be accompanied by a decrease in the amount of free energy:
\begin{equation}
\Delta{W}=-a\cdot n_s + \frac{b}{2}n_s^2,
\end{equation}
 where $a$ and $b$ are model parameters.
Using the principle of minimum free energy of the system in a steady state, we can find the relation between these parameters:
\begin{equation}
\frac{d(\Delta{W})}{dn_s}=-a+b\cdot n_s=0.
\end{equation}
Whence
 \begin{equation}
b=\frac{a}{n_s}
\end{equation}
and the energy gain in the transition to an ordered state:
\begin{equation}
\Delta W = -\frac{a}{2}n_s.
\end{equation}
The reverse transition from the superconducting state to a normal state occurs at the critical magnetic field strength, $H_c$.
This is required to create the density of the magnetic energy  $\frac{H_c^2}{8\pi}$.
According to the above description, this equation is therefore obtained:
\begin{equation}
\frac{H_c^2}{8\pi}=\frac{a}{2}n_s.
\end{equation}
In order to express the parameter  $a$ of GL-theory in terms of physical characteristics of a sample, the  density of "superconducting" \ carriers generally charge from the London's equation (\ref{lamb}).\footnote{It should be noted that due to the fact that the London equation does not correctly describes the ratio of the penetration depth with a density of carriers, one should  used  the revised equation (\ref{lam}) in order to find the $a$.}

The important step in the Ginzburg-Landau theory is the changeover of the concentration of ”superconducting” \ carriers, $n_s$, to the order parameter
 $\Psi$
\begin{equation}
 |\Psi(x)|^2={n_s}.
\end{equation}
At this the standard Schrodinger equation (in case of one dimension) takes the form:
\begin{equation}
-\frac{\hbar}{2m}\left[\nabla \Psi(x) \right]^2 - a \Psi^2(x) +\frac{b}{2}\Psi^4(x)=\mathcal{E}.\label{Shr}
\end{equation}

  Again using the condition of minimum energy
  \begin{equation}
\frac{d\mathcal{E}}{d \Psi}=0
\end{equation}
after the simple transformations one can obtain the equation that is satisfied by the order parameter of the equilibrium system:
\begin{equation}
a \Psi +b\Psi|\Psi|^2+\frac{1}{4m_e}\left(i\hbar\nabla +\frac{2e}{c}\mathbf{A}\right)^2\Psi = 0.
\end{equation}
This equation is called the first Ginzburg-Landau equation. It is nonlinear. Although there is no analytical solution for it, by using the series expansion of parameters, we can find solutions to many of the problems  which are associated with changing the order parameter. Such there are consideration of the physics of thin superconducting films, boundaries of superconductor-metal,  phenomena near the critical temperature and so on.
The variation of the Schrodinger equation (\ref{Shr}) with respect the vector potential $\mathbf{A}$ gives the second equation of the GL-theory:
\begin{equation}
\mathbf{j}_s  = \frac{i\hbar e}{2m_e}\left(\Psi^\ast\nabla\Psi - \Psi\nabla\Psi^\ast\right) -\frac{2e^2}{m_e c}|\Psi|^2\mathbf{A}.\label{js}
\end{equation}
This determines the density of superconducting current. This equation allows us to obtain a clear picture of the important effect of superconductivity: the magnetic flux quantization.

\section[Experimental data]{Experimental data that are important for creation of the theory of superconductivity}
\subsection{Features of the phase transition}
Phase transitions can occur with a jump of the first derivatives of thermodynamic potential and with a jump of second derivatives at the continuous change of the first derivatives.  In the terminology of Landau, there are two types of phase transitions: the 1st and the 2nd types.  Phenomena with rearrangement of the crystal structure of matter are considered to be a phase transition of the 1st type, while the order-disorder transitions relate to the 2nd type.
Measurements show that at the superconducting transition there are no changes in the crystal structure or the latent heat release and similar phenomena characteristic of first-order transitions.  On the contrary, the specific heat at the point of this transition is discontinuous (see below). These findings clearly indicate that the superconducting transition is associated with a change order. The complete absence of changes of the crystal lattice structure, proven by X-ray measurements, suggests that this transition occurs as an ordering in the electron subsystem.

\subsection{The energy gap and specific heat of a superconductor}

\subsubsection{\it{The energy gap of a superconductor}}

Along with the X-ray studies that show no structural changes at the superconducting transition, no changes can be seen in the optical range.  When viewing with the ‘naked eye’ here, nothing happens.  However, the reflection of radio waves undergoes a significant change in the transition.   Detailed measurements show that there is a sharp boundary in the wavelength range 1 • 1011 ÷ 5 • 1011  Hz,  which is different for different superconductors. This phenomenon clearly indicates on the existence of a threshold energy, which is needed for the transition of a superconducting carrier to normal state, i.e., there is an energy gap between these two states.

\newpage
\subsubsection{\it{The specific heat of a superconductor}}
The laws of thermodynamics  provide possibility  for an idea of the nature of the phenomena by means of general reasoning.
We show that the simple application of thermodynamic relations leads to the conclusion that the transition of a normal metal-superconductor transition is the transition of second order, i.e., it is due to the ordering of the electronic system.

In order to convert the superconductor into a normal state, we can do this via a critical magnetic field, $H_c$. This transition means that the difference between the free energy of a bulk sample (per unit of volume) in normal and superconducting states complements the energy density of the critical magnetic field:
\begin{equation}
F_n -Fs=\frac{H_c^2}{8\pi}\label{FF}.
\end{equation}
By definition, the free energy is the difference of the internal energy, $U$, and thermal energy $TS$ (where $S$ is the entropy of a state):
\begin{equation}
F=U-TS.
\end{equation}
Therefore, the increment of free energy is
\begin{equation}
\delta F = \delta U - T\delta S - S\delta T.
\end{equation}

According to the first law of thermodynamics, the increment of the density of thermal energy $\delta Q$
is the sum of the work made by a sample on external bodies
 $\delta R$, and the increment of its internal energy $\delta U$:
 \begin{equation}
\delta Q = \delta R +\delta F
\end{equation}
as a reversible process heat increment of $\delta Q =T \delta S$, then
\begin{equation}
\delta F = -\delta R -S\delta T
\end{equation}
thus the entropy
\begin{equation}
S = - \left(\frac{\partial F}{\partial T}\right)_R.
\end{equation}

In accordance with this  equation, the difference of entropy in normal and superconducting states (\ref{FF}) can be written as:
\begin{equation}
S_s - S_n = \frac{H_c}{4\pi}\left(\frac{\partial H_c}{\partial T}\right)_R\label{Ss}.
\end{equation}
Since  critical field at any temperature decreases with rising temperature:
\begin{equation}
\left(\frac{\partial H_c}{\partial T}\right)< 0,
\end{equation}
then we can conclude (from equation (\ref{Ss})),  that the superconducting state is more ordered  and therefore its entropy is lower. Besides this, since at $T=0$, the derivative of the critical field is also zero, then the entropy of the normal and superconducting state, at this point, are equal. Any abrupt changes of the first derivatives of the thermodynamic potential must also be absent, i.e., this transition is a transition of the order-disorder in electron system.

Since, by definition, the specific heat $C=T\left(\frac{\partial S}{\partial T}\right)$, then the difference of specific heats of superconducting and normal states:
\begin{equation}
C_s - C_n = \frac{T}{4\pi}\left[\left(\frac{\partial H_c}{\partial T}\right)^2 + H_c\frac{\partial^2 H_c}{\partial T^2}\right]\label{CC}.
\end{equation}
Since at the critical point $ H_c = 0 $, then from (\ref{CC}) this follows directly the Rutgers formula that determines the value of a specific heat jump at the transition point:
\begin{equation}
C_s - C_n = \frac{T}{4\pi}\left(\frac{\partial H_c}{\partial T}\right)_{T_c}^2 \label{Rut}.
\end{equation}

\vspace{0.2cm}

\begin{figure}
\hspace{1.5cm}
\includegraphics[scale=0.7]{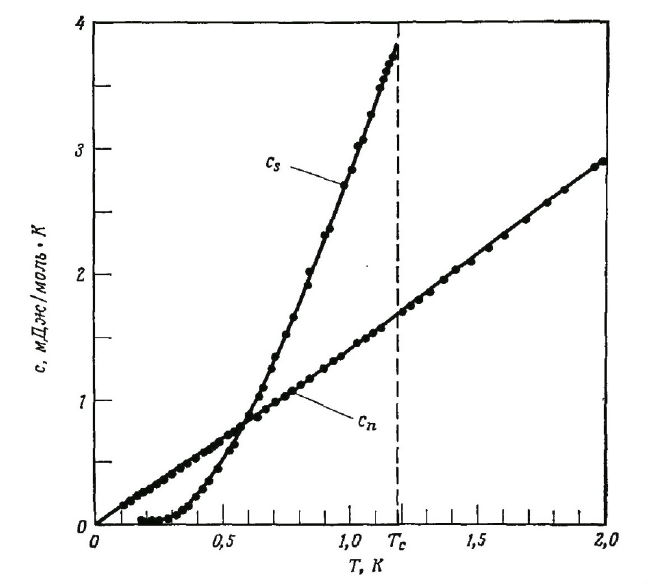}
\caption {Low-temperature heat capacity of normal and superconducting aluminum\cite{heat}.}\label{heat}
\end{figure}

\vspace{0.2cm}

The theory of the specific heat of superconductors is well-confirmed experimentally.
For example, the low-temperature specific heat of aluminum in both the superconducting and normal states supports this in Fig.(\ref{heat}).
Only the electrons determine the heat capacity of the normal aluminum at this temperature, and in accordance with the theory of Sommerfeld, it is linear in temperature. The specific heat of a superconductor at a low temperature is exponentially dependent on it.  This indicates the existence of a two-tier system in the energy distribution of the superconducting particles. The measurements of the specific heat jump at
$T_c $ is well described by the Rutgers equation(\ref{Rut}).

\newpage

\subsection{Magnetic flux quantization in superconductors}\label{qu}
The conclusion that magnetic flux in hollow superconducting cylinders should be quantized was firstly expressed F.London. However, the main interest in this problem is not in the phenomenon of quantization, but in the details: what should be the value of the flux quantum. F.London had not taken into account the effect of coupling of superconducting carriers when he computed the quantum of magnetic flux and, therefore, predicted for it twice the amount.
The order parameter can be written as:
\begin{equation}
\Psi(r)=\sqrt{n_s}~e^{i\theta(r)}\label{Psi}.
\end{equation}
where $n_s$ is density of superconducting carriers, $\theta$ is the order parameter phase.

As in the absence of a magnetic field, the density of particle flux is described by the equation:
\begin{equation}
n_s\mathbf{v}=\frac{i\hbar}{2m_e}\left(\Psi\nabla\Psi^\ast-\Psi^\ast\nabla\Psi\right). 
\end{equation}
Using (\ref{Psi}), we get $\hbar\mathbf{\nabla}\theta=2m_e\mathbf{v_s}$ and can transform the Ginzburg-Landau equation (\ref{js}) to the form:
\begin{equation}
\hbar\mathbf{\nabla}\theta = 2m_e \mathbf{v}_s + \frac{2e}{c}\mathbf{A} \label{F00}.
\end{equation}
If we consider the freezing of magnetic flux in a thick superconducting ring with a wall  which thickness is much larger than the London penetration depth $\lambda_L$, in the depths of the body of the ring current density of $\mathbf{j}$ is zero. This means that the equation (\ref{F00}) reduces to the equation:
\begin{equation}
\hbar\mathbf{\nabla}\theta =  \frac{2e}{c}\mathbf{A} \label{F01}.
\end{equation}
One can take the integrals on a path that passes in the interior of the ring, not going close to its surface at any point,  on the variables included in this equation:
\begin{equation}
\hbar\oint\mathbf{\nabla}\theta ds=  \frac{2e}{c}\oint\mathbf{A}ds \label{F02},
\end{equation}
and obtain
\begin{equation}
\oint\mathbf{\nabla}\theta ds=  \frac{2e}{\hbar c}\Phi,
\end{equation}
since by definition, the magnetic flux through any loop:
\begin{equation}
\Phi=\oint\mathbf{A}ds. 
\end{equation}
The contour integral $\oint\mathbf{\nabla}\theta ds$ must be a multiple of $2\pi$, to ensure the uniqueness of the order parameter in a circuit along the path.
 Thus, the magnetic flux trapped by superconducting ring should be a multiple to  the quantum of magnetic flux:
\begin{equation}
\Phi_0=\frac{2\pi \hbar c}{2e}, 
\end{equation}
that  is confirmed by corresponding measurements.

 This is a very important result for understanding the physics of superconductivity. Thus, the theoretical predictions are confirmed by measurements saying that the superconductivity is due to the fact that its carriers have charge 2e, i.e., they represent two paired electrons. It should be noted that the pairing of electrons is a necessary condition for the existence of superconductivity,  but this phenomenon was observed  experimentally in the normal state of electron gas metal too. The value of the quantum Eq.(36)  correctly describes the periodicity of the magnetic field influence on electron gas in the normal state of some metals (for example, Mg and Al at temperatures much higher than their critical temperatures)\cite{Shab}, \cite{Sharv}).

\subsection{The isotope effect}
The most important yet negative role, which plays a major part in the development of the science of superconductivity, is the isotope effect, which was discovered in 1950. The negative role, of the isotope effect is played not just by the effect itself but its wrong interpretation.
It was established through an experiment that the critical temperatures of superconductors depend on the isotope mass Mi, from which they are made:
\begin{equation}
T_c\sim \frac{1}{{M_i}^{a}}\label{is}.
\end{equation}
This dependence was called the isotope effect.
It was found that for type-I superconductors - $Zn, Sn, In, Hg, Pb$ - the  value of the isotope effect can be described by Eq.(\ref{is}) at the constant $a=\frac{1}{2}$.

This effect has made researchers think that the phenomenon of superconductivity is actually associated with the vibrations of ions in the lattice. This is because of the similar dependence (Eq.(\ref{is})) on the ion mass in order for the maximum energy of phonons to exist whilst propagating in the lattice.

Subsequently this simple picture was broken: the isotope effect was measured for other metals, and it had different values. This difference of the isotope effect in different superconductors could not be explained  by phonon mechanism.

\begin{figure}
\hspace{1cm}
\includegraphics[scale=0.5]{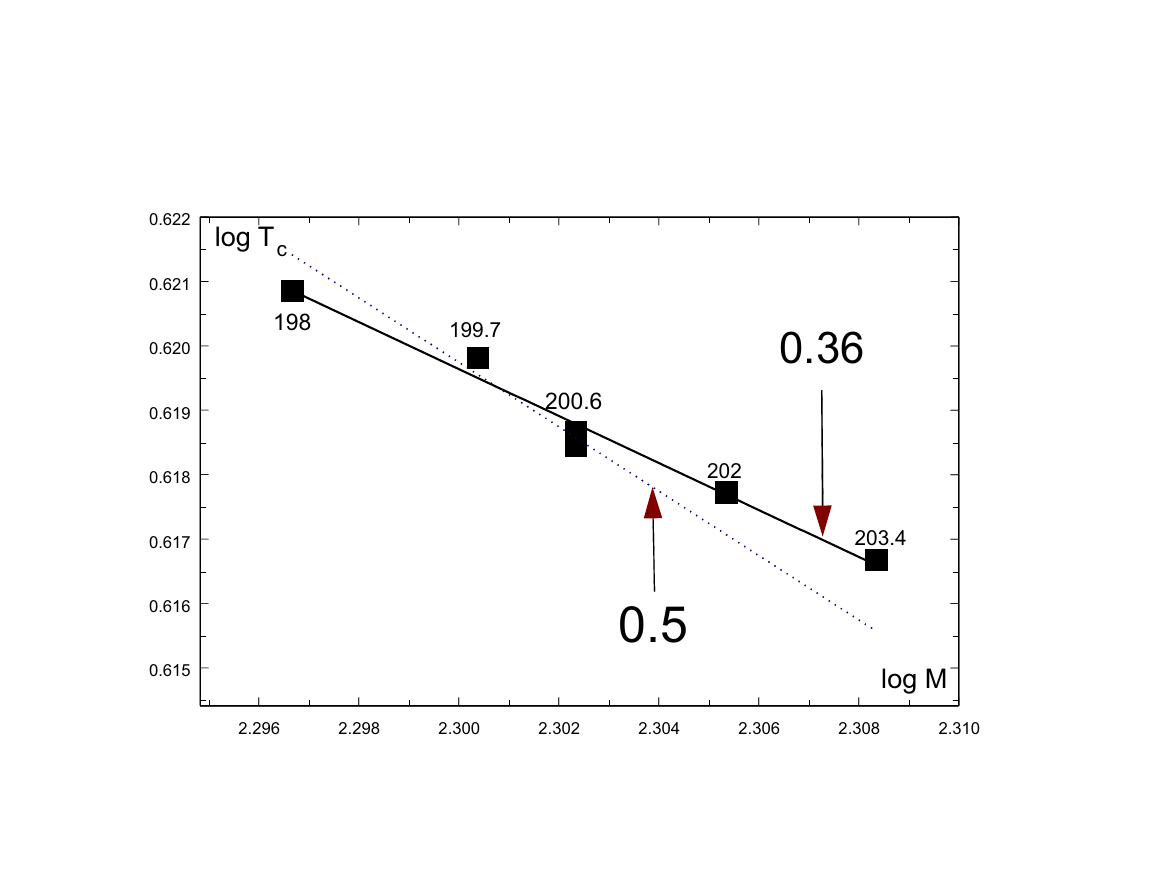}
\caption {The isotope effect in mercury.
The solid line is obtained by the sparse-squares technique.
In accordance with the phonon mechanism, the coefficient  $\mathfrak{a}$ must be about 1/2 (the dotted line).
As it can be seen, this coefficient is in reality  approximately equal to  1/3.}\label{Hgg}
\end{figure}

It should be noted that the interpretation of the isotope effect in the ”simple” \ metals did exist though it seemed to fit the results of measurements under the effect of phonons, where $a=1/2$. Since the analysis of experimental data \cite{Maxwell}, \cite{Serin} (see Fig.(\ref{Hgg})) suggests that this parameter for mercury is really closer to 1/3.

\newpage

\section{BCS}
\begin{figure}
\includegraphics[scale=1]{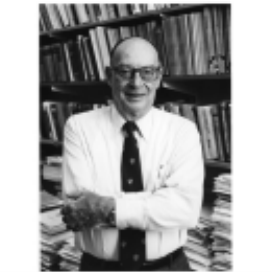}
\caption{John Bardeen - twice winner of the Nobel Prize.
He received the Prize in 1956 for the invention of the transistor,
and  in 1972, along with L.Cooper  and J.Shrieffer, for the creation of the BCS-theory.
{\it  I would like to put here somewhere my photo at the conversation  with John Bardeen.
It would be a remarkable illustration of the continuity of generations within the science of superconductivity!
The great thing that this photo must exist.
In the late 1980s, I was at the conference on superconductivity at Stanford University. I met J. Bardeen, who also attended the conference. It was there that I spoke to him. While we were talking, I saw an American physicist taking a photo of us. I knew this physicist at the time, because he had attend my laboratory in Dubna to acquaint himself with the work of high-Tc SQUID. Our laboratory gained international recognition, because we were the first scientific team in the world who could overcome the natural barrier of SQUID sensitivity \cite{Lik}. This work was aimed at measuring the magnetic cardiogram of humans with the help of high-Tc SQUID.
 After twenty-odd years now though, I cannot remember neither the name of this American scientist, nor even where he had come from. I hope the photographs taken during his visit to Dubna still exist. So there is still hope that with the help of American friends I can find him, and with him, those photos made more than twenty years ago.}}
 \label{bardg}
\end{figure}

The first attempt to detect the isotope effect in lead was made by the Leiden group in the early 1920s, but due to a smallness of the effect it was not found. It was then registered in 1950 by researchers of the two different laboratories. It created the impression that phonons are responsible for the occurrence of superconductivity since the critical parameters of the superconductor depends on the ion mass. In the same year H.Fr\"{o}hlich was the first to point out that at low temperatures, the interaction with phonons can lead to nascency of forces of attraction between the electrons, in spite of the Coulomb repulsion. A few years later, L. Cooper predicted the specific mechanism in which an arbitrarily weak attraction between electrons with the Fermi energy would lead to the emergence of a bound state. On this basis, in 1956, Bardeen, Cooper and Shrieffer built a microscopic theory, based on the electron-phonon interaction as the cause of the attractive forces between electrons.\\

It is believed that the BCS-theory has the following main results:\\

1. The attraction in the electron system arises due to the electron-phonon interaction. As result of this attraction, the ground state of the electron system is separated from the excited electrons by an energetic gap. The existence of energetic gap explains the behavior of the specific heat of superconductors, optical experiments and so on.\\

2. The depth of penetration (as well as the coherence length) appears to be a natural consequence of the ground state of the BCS-theory. The London equations and the Meissner diamagnetism are obtained naturally.\\

3. The criterion for the existence of superconductivity and the critical temperature of the transition involves itself the density of electronic states at the Fermi level $\mathcal{D}(\mathcal{E}_F)$ and the potential of the electron-lattice interaction $U$, which can be estimated from the electrical resistance. In the case of $U\mathcal{D}(\mathcal{E}_F)\ll 1$ the BCS-theory expresses the critical temperature of the superconductor  in terms of its Debye temperature $\Theta_D$:
\begin{equation}
T_c=1.14\cdot \Theta_D \cdot exp\left[-\frac{1}{U\mathcal{D}(\mathcal{E}_F)}\right].\label{dbk}
\end{equation}\\

4. The participation of the lattice in the electron-electron association determines the effect of isotopic substitution on the critical temperature of the superconductor. At the same time due to the fact that the mass of the isotopes depends on the Debye temperature $\theta_D\propto M^{-1/2}$, Eq.(\ref{dbk}) correctly describes this relationship for a number of superconductors.\\

\begin{figure}
\hspace{1.5cm}
\includegraphics[scale=0.5]{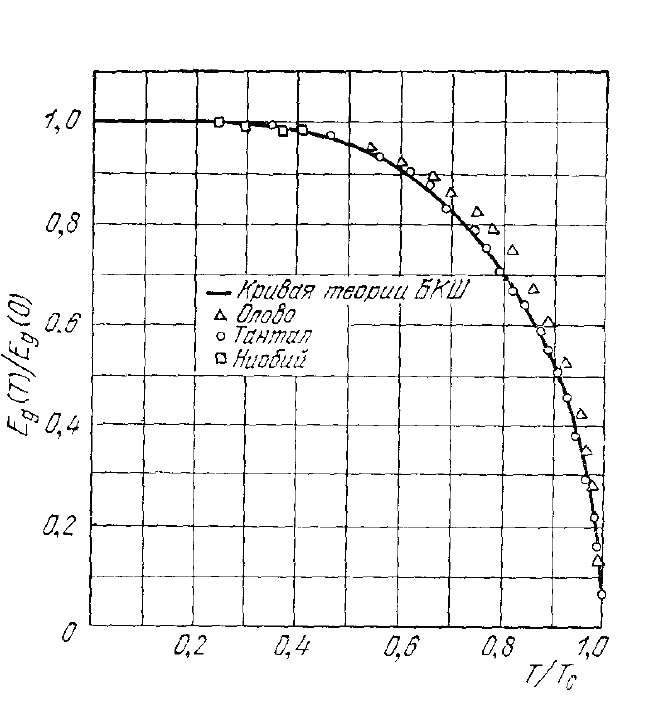}
\caption{The temperature dependence of the gap in the energy spectrum of superconductors, calculated by the Eq.(\ref{1d}).}\label{1de}
\end{figure}

5. The temperature dependence of the energy gap ∆ of the superconductor is described in the BCS-theory implicitly by an integral over the phonon spectrum from 0 to the Debye energy:
\begin{equation}
1=\frac{U\mathcal{D}(\mathcal{E}_F)}{2}\int_0^{\hbar\omega_D} d\xi\frac{th\sqrt{\xi^2+\Delta^2}/2kT}{\sqrt{\xi^2+\Delta^2}}.\label{1d}
\end{equation}
The result of calculation of this dependence is in good agreement with measured data(Fig.(\ref{1de})).\\
\\

6. The BCS-theory is consistent with the data of measurements of the magnetic flux quantum, as its ground state is made by pairs of single-electron states.\\

But all this is not in a good agreement with this theory.\\

 First, it does not give the main answer to our questions: with using of it one cannot obtain apriori information about what the critical parameters of a particular superconductor should be. Therefore, BCS cannot help in the search for strategic development of superconductors or in the tactics of their research. Eq.(\ref{dbk}) contains two parameters that are difficult to assess: the value of the electron-phonon interaction and the density of electronic levels near the Fermi level. Therefore, BCS cannot be used solely for this purpose, and can only give a qualitative assessment.

In addition, many results of the BCS theory can be obtained by using other, simpler but `fully correct` methods.

The coupling of electrons in pairs can be the result not only of electron-phonon mechanism.
Any attraction between the electrons can lead to their coupling.

For the existence of superconductivity, the bond energy should combine into single ensembles of separate pairs of electrons, which are located at distances of approximately hundreds of interatomic distances. In BCS-theory, there are no forces of attraction between the pairs and, especially, between pairs on these distances.

The quantization of flux is well described within the Ginzburg-Landau theory (see Sec.(\ref{qu})),  if the order parameter describes the density of paired electrons.

By using a two-tier system with the approximate parameters, it is easy to describe the temperature  dependence of the specific heat of superconductors.

So, the calculation of the temperature dependence of the superconducting energy gap formula (Eq.(\ref{1d})) can be considered as the success of the BCS theory .

However, it is easier and more convenient to describe this phenomenon as a characterization of the order-disorder transition in a two-tier system of zero-point fluctuations of condensate.
In this approach, which is discussed below in Sec.(\ref{crit-param1}), the temperature dependence of the energy gap receives the same interpretation as other phenomena of the same class — such as the $\lambda$-transition in liquid helium, the temperature dependence of spontaneous magnetization of ferromagnets and so on.\\

Therefore, as in the 1950s, the existence of isotope effect is seen to be crucial.
However, to date, there is experimental evidence that shows the isotope substitution leads to a change of the parameters of the metals crystal lattice due to the influence of isotope mass on the zero-point oscillations of the ions (see \cite{Inyu}).
For this reason, the isotope substitution in the lattice of the metal should lead to a change in the Fermi energy and its impact on all of its electronic properties. In connection with this, the changing of the critical temperature of the superconductor at the isotope substitution can be a direct consequence of changing the Fermi energy without any participation of phonons.\\

The second part of this book will be devoted to the role of the ordering of zero-point oscillations of electrons in the mechanism of the superconducting state formation.

\newpage

\newpage

\section{The new Era -  HTSC}
\label{HT}
During the century following the discovery of superconductivity, 40 pure metals were observed. It turned out that among them, Magnesium has the lowest critical temperature — of about 0.001K, and Technetium has the highest at 11.3K.

 Also, it was found that hundreds of compounds and alloys at low temperatures have the property of superconductivity. Among them, the intermetallic compound $Nb_3 Ge$ has the highest critical  temperature — 23.2K.

 In order to obtain the superconducting state in these compounds it is necessary to use the expensive technology of liquid helium.\footnote{For comparison, we can say that liter of liquid helium costs about a price of bottle of a good brandy, and the heat of vaporization of helium is so small that expensive  cryostats are needed for its storage. That makes its using very expensive.}

Theoretically, it seems that liquid hydrogen could also be used in some cases. But this point is still more theoretical consideration than practical one: hydrogen is a very explosive substance.
For decades scientists have nurtured a dream to create a superconductor which would retain its properties at temperatures  above the boiling point of liquid nitrogen. \\

\begin{figure}
\includegraphics[scale=1.2]{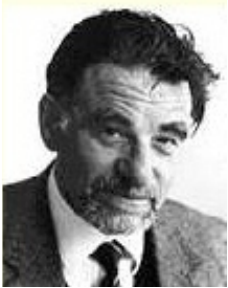}
\caption {Karl Alexander Muller - founder of HTSC.}
\end{figure}
\label{alexg}

Liquid nitrogen is cheap, accessible, safe, and a subject to a certain culture of working with him (or at least it is not explosive). The creation of such superconductor promised breakthrough in many areas of technology.\\

In 1986, these materials were found. At first, Swiss researchers  A.Muller and G.Bednortz found the superconductivity in the copper-lanthanum ceramics, which temperature of superconducting  transition was ”only” \ 40K, and soon it became clear that it was the new class of superconductors (they was called high-temperature superconductors, or HTSC), and a very large number of laboratories around the world were included in studies of these materials.

\vspace{0.5cm}
\begin{table}
\centering
\begin{tabular}{||c|c|c||}\hline\hline
  superconductor&$T_c,K$&$H_c,Oe$\\\hline
  $Hg$                     &  4.15     &  41 \\
  $Pb$                     &  7.2      &  80 \\
  $Nb$                     &  9.25     & 206 \\
  $NbTi$                   & 9.5-10.5  &120.000\\
  $Nb_3Sn$                 & 18.1-18.5 &220.000\\
  $Nb_3Al$                 & 18.9      &300.000\\
  $Nb_3Ge$                 & 23.2      &370.000\\
  $MgB_2$                  &~40        &150.000\\
  $YBa_2Cu_3O_7$           & 92.4      &600.000\\
  $Bi_2Sr_2Ca_2Cu_3O_{10}$ & 111       &~5.000.000\\
  $HgBa_2Ca_2Cu_3O_8$      & 133       &$>$10.000.000\\\hline\hline
\end{tabular}
\caption{Critical parameters of superconductors}\label{su}
\end{table}
\bigskip

One year later, the superconductivity was discovered in ceramic $ Y − Ba − C u − O$ with transition temperature higher than 90K.  As the liquid nitrogen boils at 78K, the nitrogen level was overcome.

\begin{figure}
\includegraphics[scale=.5]{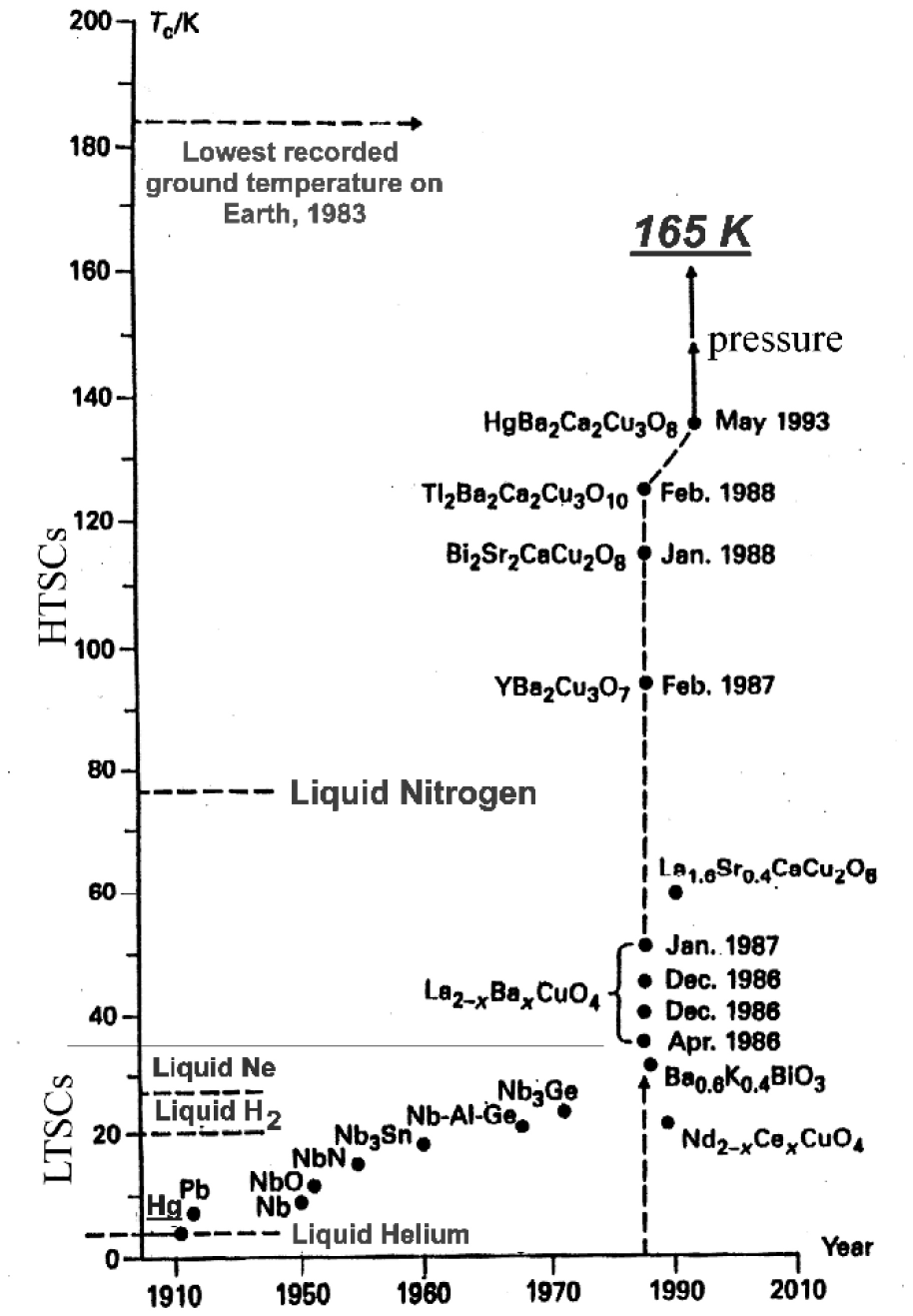}
\caption{Growth of the critical temperature of known superconductors in time.
The horizontal lines indicate the temperature of boiling of cryogenic liquids (at normal pressure).}
\label{htscg}
\end{figure}

Soon after it, mercury based ceramics with transition temperatures of the order of 140K were found.  The history of increasing the critical temperature of the superconductor is interesting to trace: see the graph (\ref{htscg}).


It is clear from this graph that if the creation of new superconductors has been continuing at the same rate as before the discovery of HTSC, the nitrogen levels would have been overcome through 150 years. But science is developing by its own laws, and the discovery of HTSC allowed to raise sharply the critical temperature.\\

However, the creation of high-Tc superconductors has not led to a revolutionary breakthrough in technology and industry. Ceramics were too low-technological to manufacture a thin superconducting wires. Without wires, the using of high-Tc superconductors had to be limited by low-current instrument technology. It also did not cause the big breakthroughs in this manner, either (see, e.g. \cite{Lik}).

After discoveries of high-Tc superconductors, no new fundamental breakthrough of this value  was made.
Perhaps the reason for this is that the BCS-theory, adopted by most researchers to date, can not predict the parameters of superconductors apriori and serves as just a support in strategy and tactics of their research.

\newpage

\chapter{Superfuidity}
\begin{figure}
\hspace{3cm}
\includegraphics[scale=0.5]{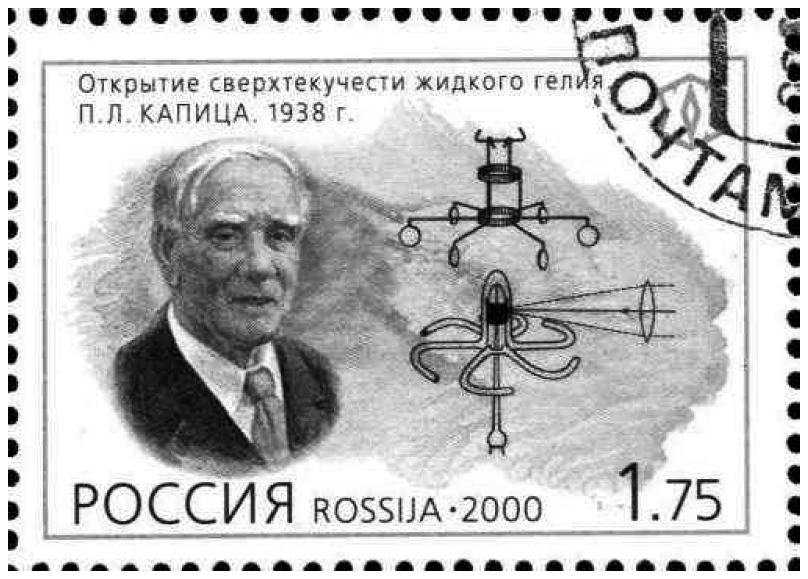}
\vspace{-7cm}\caption{Pyotr L.Kapitsa}
\end{figure}
\label{kapg}

The first study of the properties of helium-II began in the Leiden laboratory as early as in 1911, the same year when superconductivity was discovered.  A little later the singularity in the specific heat, called the $\lambda$-transition, was discovered.  However, the discovery of superfluidity of liquid helium was still far away, as Pyotr L. Kapitsa discovered it by 1938.\\

{\it{This discovery became a landmark for the world science, so many events associated with it are widely known, but one story of Kapitsa`s background was never published.\\

This Soviet scientist Isaak K. Kikoine relayed this story to me in the mid 1960s, at this time, when I was his graduate student. Kikoine was one of the leaders of the Soviet atomic project and was also  engaged with important state affairs for most of the days. In the evenings, however, he often visited the labs of my colleagues or my laboratory to discuss scientific news. During these talks, he often intertwined scientific debate and interesting memories from history of physics.\\

Here is his story about Kapitsa and superfluidity as it was remembered and relayed to me.\\
\begin{figure}
\hspace{3cm}
\includegraphics[scale=0.7]{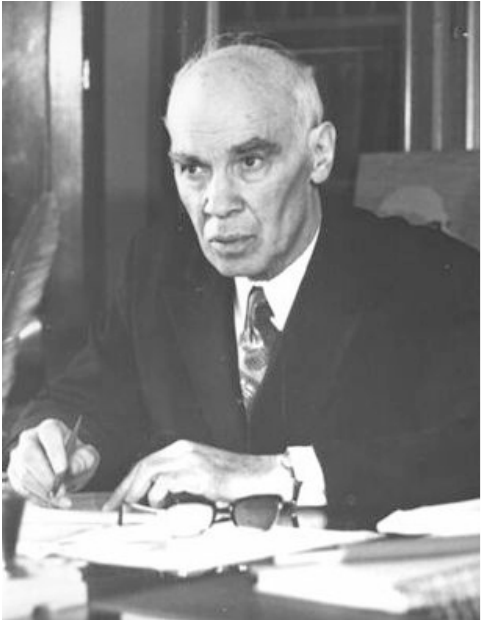}
\caption {Isaak K.Kikoine}
\end{figure}
\label{ikkg}

It happened in 1933 when Isaac Kikoine was 25 years old. He had just completed his experiment of measuring
of the gyromagnetic effect in superconductors.  Pyotr Kaptsa was aware of this  experiment, he even sent a small ball of super-pure lead for this experiment all the way from the Mond Laboratory in Cambridge, which he led then.

Almost every summer, he went with his family on his own car to Crimea.  This was an absolute luxury for Soviet people at that time! \\
On his way he visited  physical laboratories in Moscow and Leningrad (now again St.  Petersburg), lecturing and networking with colleagues, friends and admirers.
During one of these visits, in 1933, Kikoine had a chance to talk to Kapitsa about results of his measurements.  Kapitsa liked these results, and he invited Kikoine to work in Cambridge. \\
They had arranged all formalities, including that Kapitsa will send him an invitation to work in England must be organized for the year.\\
They planned to go to Cambridge together after the next summer vacation.\\
But it has not happen. \\
Summer 1934 Kapitsa as usual came to Russia. But when he wanted to go back  to England, his return visa was canceled. No efforts helped him as it had been decided by the authorities at the top.\\
Kapitsa`s father-in-law was  A.N. Krylov was a famous ship-builder.
 Krylov together with his friend Ivan Pavlov (Nobel Laureate in physiology from pre-revolutionary  times) asked for an audience with Stalin himself.
  Stalin did receive them and asked them:\\
  - What is the problem? \\
  - Yes—oh, we are asking you to allow Kapitsa to go  abroad.\\
  - He will not be released. Because the Russian nightingale must sing in Russia!”\\
Pavlov (physiologist) said in reply:\\
- With all respect, a nightingale does not sing in a cage!\\
- Anyway he would sing for us!\\

So terribly unkind (it is a some understatement yet) Kapitsa became the deputy director of the Leningrad Physical-Technical Institute, which was directed by A.F. Ioffe. \\
Young professors of this institution - Kikoine,  Kurchatov, Alikhanov, Artsimovich then went to see the Deputy Director. \\
Alikhanov asked at the door of the office from outgoing Artsimovich: \\
- Well, how is he? A beast or a man?\\
- A centaur! - was answer.\\

This nickname stuck firmly to P. Kapitsa. \\
The scientists of the older generation called him by the same nickname, even decades later.

Stalin, of course, was a criminal.
Thanks to his efforts,  almost each family in the vast country lost some of its members in the terror of unjustified executions or imprisonments.\\
However Stalin`s role in the history of superfluidity can be considered  positive.\\
 All was going  well for Kapitsa in Russia,\\
if to consider researches of  superfluidity.}}\\

\newpage

 \begin{figure}
\hspace{2cm}
\includegraphics[scale=1]{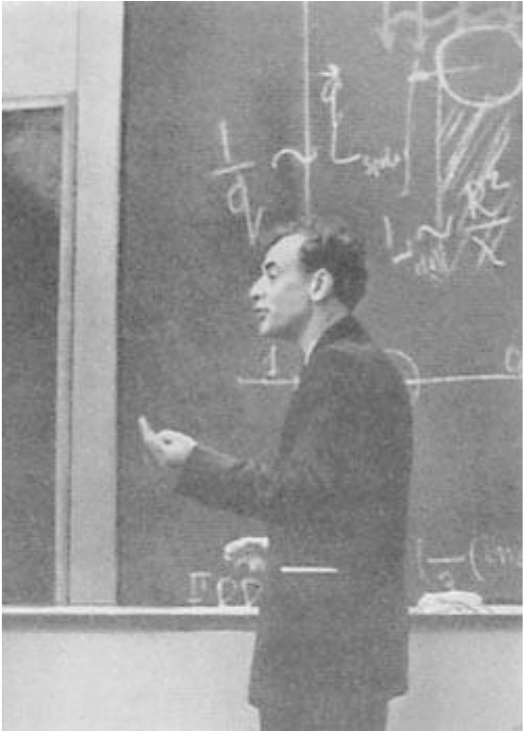}
\caption {L.D. Landau}
\end{figure}
\label{daug}

Already after a couple of years, L.Landau\footnote{Before P.Kapitsa spent a lot of the courage to pull out L.Landau from the stalinist prison.} was able to give a theoretical explanation of this phenomenon.\\
He viewed helium-II as a substance where the laws of quantum physics worked on a macroscopic scale.\\

This phenomenon is akin to superconductivity: the superconductivity can be regarded as the superfluidity of an electron liquid.
As a result, the relationship between phenomena have much in common, since both phenomena  are described by the same quantum mechanics laws in macroscopic manifestations.  This alliance, which exists in the nature of the phenomena, as a consequence, manifests itself in a set of physical phenomena: the same laws of quantum effects  and the same physical description.

For example, even the subtle quantum effect of superconductors, such as the tunneling Josephson effect, manifests itself in the case of superfluidity as well.\\

However, there are some differences.\\
At zero temperature, only a small number of all conduction electrons form the superfluid component in superconductors.

 The concentration of this component is of the order $\frac{kT_c}{\mathcal{E}_F}\approx 10^{-5}$, while in liquid helium at $T=0 $, all liquid becomes superfluid, i.e., the concentration of the superfluid component is equal to 1.

It is also significant to note that the electrons which have an interaction with their environment have an electric charge and magnetic moment. This is the case yet it seems at first glance that there are no mechanisms of interaction for the formation of the superfluid condensate in liquid helium. Since the helium atom is electrically neutral, it has neither spin, nor magnetic moment.

By studying the properties of helium-II, it seems that all main aspects of the superfluidity has been considered. These include: the calculations of density of the superfluid component and its temperature dependence, the critical velocity of superfluid environment and its sound, the behavior of superfluid liquid near a solid wall and near the critical temperature and so on. These issues, as well as some others, are considered significantly in many high-quality original papers and reviews \cite{Halat} - \cite{Vol}.
 There is no need to rewrite their content here.

 However, the electromagnetic mechanism of transition to the superfluid state in helium-4 still remains unclear, as it takes place at a temperature of about one Kelvin and also in the case of helium-3, at the temperature of about a thousand times smaller.
It is obvious that this mechanism should be electromagnetic.

This is evidenced by the scale of the energy at which it occurs. The possible mechanism for the formation of superfluidity will be briefly discussed in the Chapter (\ref{HeE}).


\part{Superconductivity,\\
 superfluidity\\
 and zero-point oscillations}
{
{
\hspace{4cm} Instead an epigraph\\

{\it{

There is a  parable  that the population of one small south russian town in the old days  was divided between  parishioners of the Christian church and the Jewish synagogue.

A Christian priest was already heavily wiser by a life experience, and the rabbi was still quite young and energetic.

 One day the rabbi came to the priest for advice.\\
- My colleague, - he said, - I lost my bike. I feel that it was stolen by  someone from there, but who did it I can not identify. Tell me what to do.\\
-Yes, I know one scientific method for this case, - replied the priest.

You should do this: invite all your parishioners  the synagogue and read them the $\ll$Ten Commandments of Moses$\gg$.
When you read $\ll$Thou shalt not steal$\gg$, lift your head and look carefully into the eyes of your listeners. The listener who turns his eyes aside will be the guilty party.

A few days later, the rabbi comes to visit the priest with a bottle of Easter-vodka and on his bike. \\
The priest asked the rabbi to tell him what happened in details.\\
The rabbi told the priest that his theory had worked in practice and the bike was found.
The rabbi relayed the story:

 - I collected my parishioners and began to preach. While I approached reading the $\ll$Do not commit adultery$\gg$ commandment I remembered then where I forgotten my bike!\\

\hspace{1cm} So it is true: there is indeed nothing more practical than a good theory!\\
}}
}}
\chapter[Superconductivity and zero-point oscillations]{Superconductivity as a consequence of ordering of zero-point oscillations in electron gas}

\section[Superconductivity and zero-point oscillations]{Superconductivity as a consequence of ordering of zero-point oscillations}
Superfluidity and superconductivity, which can be regarded as the superfluidity of the electron gas, are related phenomena.
The main feature of these phenomena can be seen in a fact that a special condensate  in superconductors as well as in superfluid helium   is formed from particles  interconnected by attraction.
This mutual attraction does not allow a scattering of individual particles  on defects and walls, if the energy of this scattering is less than the energy of attraction.
Due to the lack of scattering, the condensate acquires ability to move without friction.

Superconductivity was discovered over a century ago, and the superfluidity  about thirty years later.

However, despite the attention of many scientists to the study of these phenomena, they have been the great mysteries in condensed matter physics  for a long time.
This mystery attracted  the best minds of the twentieth century.

The mystery of the superconductivity phenomenon   has begun to drop in the middle of the last century when  the effect of magnetic flux quantization in superconducting cylinders was discovered and investigated.
This phenomenon was predicted even before the WWII by brothers F. London and H. London, but its quantitative study were  performed only two decades later.

By these measurements it became clear that at the formation of the superconducting state, two free electrons are combined into a single boson with zero spin and zero momentum.

Around the same time,  it was observed that the substitution of one isotope of the superconducting element to another leads to a changing of the critical temperature of superconductors: the phenomenon  called an isotope-effect \cite{Maxwell}, \cite{Serin}. This effect was interpreted as the direct proof of the key role of phonons in the formation of the superconducting state.

Following these understandings, L. Cooper proposed the phonon  mechanism of electron pairing on which base the microscopic theory of superconductivity (so called BCS-theory) was built by N. Bogolyubov and J. Bardin, L. Cooper and  J. Shriffer (probably it should be named better the Bogolyubov-BCS-theory).

However the B-BCS theory based on the phonon mechanism brokes a hypothetic link between superconductivity and superfluidity as in liquid helium there are no phonons for combining atoms.

Something similar happened with the description of superfluidity.

Soon after discovery of superfluidity, L.D. Landau in his first papers on the subject immediately demonstrated that this superfluidity should be considered as a result of condensate formation  consisting of macroscopic number of atoms in the same quantum state and obeying quantum laws.
It gave the possibility  to describe the main features of this phenomenon: the temperature dependence of the superfluid phase density, the existence of the second sound, etc.
But it does not gave an answer to the question which physical mechanism leads to the unification of the atoms in the superfluid condensate and what is the critical temperature of the condensate, i.e. why the ratio of the temperature of transition to the superfluid state to the boiling point of helium-4 is almost exactly equals to $1/2$, while for helium-3, it is about a thousand times smaller.

On the whole, the description of both super-phenomena, superconductivity and superfluidity, to the beginning of the twenty first century induced some feeling of dissatisfaction primarily due to the fact that  a common mechanism of their occurrence has not been understood.

More than fifty years of a study  of the B-BCS-theory has shown that this theory successfully describes the general features of the phenomenon, but it can not be
developed in the theory of superconductors.
It explains general laws such as the emergence of the energy gap, the behavior of  specific heat capacity, the flux quantization, etc.,
but it can not predict the main parameters of the individual superconductors: their critical temperatures and critical magnetic fields.
More precisely, in the B-BCS-theory, the expression for the critical temperature of superconductor obtains an exponential form which exponential factor is impossible to measure directly and  this formula is of no practical interest.

Recent studies of the isotopic substitution showed that zero-point oscillations of the ions in the metal lattice are not harmonic. Consequently the isotopic substitution affects the interatomic distances in a lattice, and as the result, they directly change  the Fermi energy of a metal  \cite{Inyu}.

 Therefore, the assumption developed in
the middle of the last century, that the electron-phonon interaction is the only possible mechanism of superconductivity was  proved to be wrong.
 The direct effect of isotopic substitution on the Fermi energy gives a possibility to  consider the superconductivity without the phonon mechanism.

Furthermore, a closer look  at the problem reveals that the B-BCS-theory describes the mechanism of electron pairing, but in this theory there is no mechanism for combining pairs in the single super-ensemble.
The necessary condition for the existence of superconductivity is  formation of a unique ensemble of particles.
By this mechanism, a very small amount of electrons are combined in super-ensemble, on the level 10 in minus fifth power from the full number of free electrons.
This fact also can not be understood in the framework of the B-BCS theory.

An operation of the mechanism of electron pairing and turning them into boson pairs is a necessary but not sufficient condition for the existence of a superconducting state.
Obtained pairs are not identical at any such mechanism. They differ because of their uncorrelated zero-point oscillations and they can not form the condensate at that.

At very low temperatures, that allow superfluidity in helium and superconductivity in metals,  all movements of particles are freezed except for their zero-point oscillations.
Therefore, as an alternative, we should consider the interaction of super-particles through electro-magnetic fields of zero-point oscillations.
This approach was proved to be fruitful.
At the consideration of super-phenomena as  consequences of the zero-point oscillations ordering, one can construct theoretical mechanisms  enabling  to give  estimations for the critical parameters of these  phenomena which are in satisfactory agreement with measurements.

As result, one can see that as the critical temperatures of (type-I) superconductors are equal to about $10^{-6}$ from
the Fermi temperature for superconducting metal, which is consistent with data of measurements.
At this the destruction of superconductivity by application of critical magnetic field occurs when the field destroys the coherence of zero-point oscillations of electron pairs. This is  in good agreement with measurements also.

A such-like  mechanism works in superfluid liquid helium.
The problem of the interaction of zero-point oscillations of the electronic shells of neutral atoms in the s-state, was considered yet before the World War II by F.London.
He has shown that this interaction is responsible for the liquefaction of helium.
The closer analysis of  interactions of zero-point oscillations  for helium atomic shells shows that at first at the temperature of about 4K only, one of the oscillations mode  becomes ordered. As a result, the forces of attraction appear between  atoms which are need for helium liquefaction.
To create a single quantum ensemble, it is necessary to reach the complete ordering of atomic oscillations.
At  the complete ordering of oscillations at about 2K, the additional energy of the mutual attraction appears and the system of helium-4 atoms transits in superfluid state.
To form the superfluid quantum ensemble in Helium-3, not only  the zero-point oscillations should be ordered, but   the magnetic moments of the nuclei should be ordered too.
For this reason, it is necessary to lower the temperature below 0.001K. This is also in agreement with experiment.

Thus it is possible to show that  both related super-phenomena, superconductivity and superfluidity, are based on the single physical mechanism: the ordering of  zero-point oscillations.

The roles of zero-point oscillations in formation of the superconducting state
have been previously considered in papers \cite{BV1}-\cite{BV3}.

\section{The electron pairing}
J.Bardeen was first who turned his attention toward a possible link between superconductivity and  zero-point oscillations  \cite{Bar}.

The special role of zero-point vibrations exists due to the fact that at low temperatures  all movements of electrons in metals have been  frozen except for these oscillations.

\begin{figure}
\hspace{-3cm}
\vspace{3cm}
\includegraphics[scale=0.7]{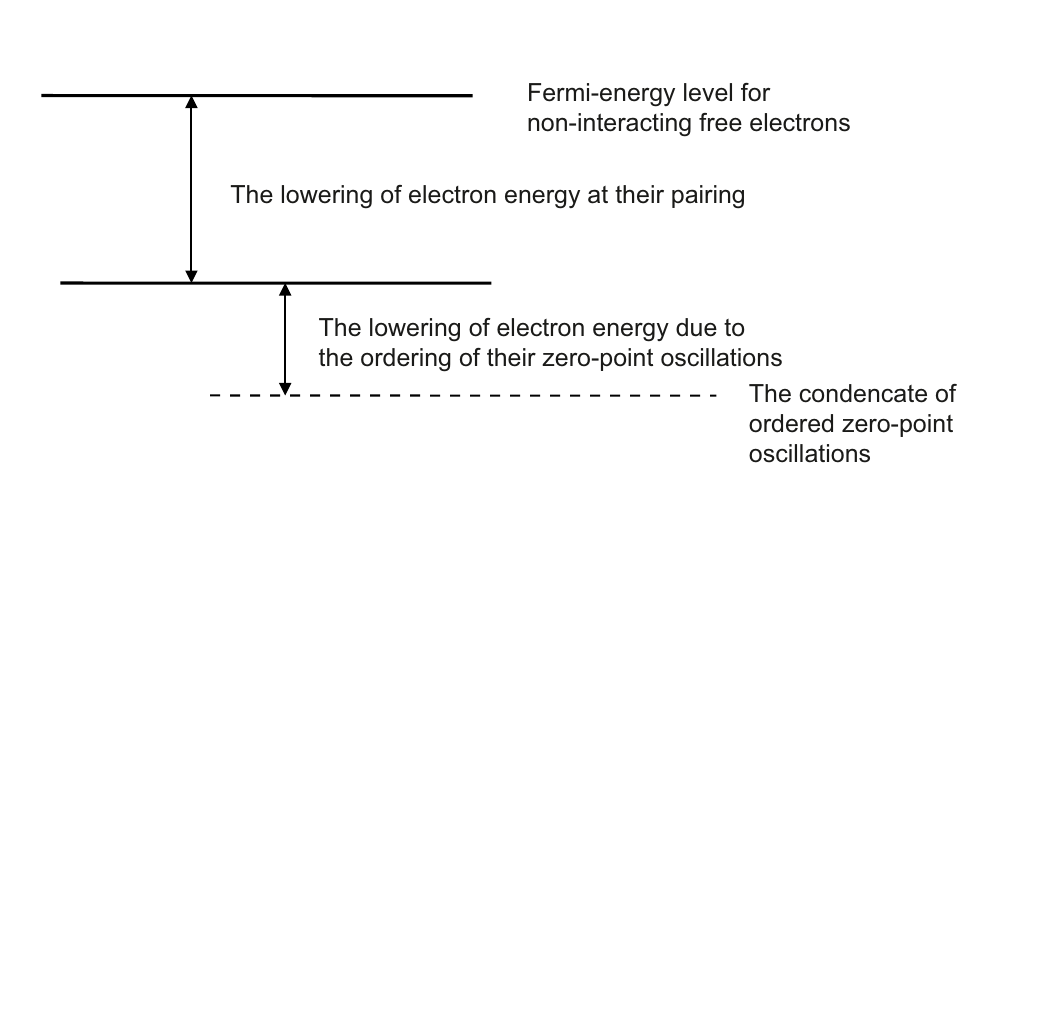}
\vspace{-8cm}\caption {The schematic representation of the energy levels of conducting electrons in a superconducting metal}\label{spe}
\end{figure}

Superconducting condensate formation requires two mechanisms:
 first, the electrons must be united in boson pairs, and then the zero-point fluctuations must be ordered (see Fig.(\ref{spe})).

The energetically favorable pairing of electrons in the electron gas should occur above the critical temperature.

Possibly, the pairing of electrons can occur due to the magnetic dipole-dipole interaction.

 For the magnetic dipole-dipole interaction, to merge two electrons  into the singlet pair at the temperature of about 10K, the distance between these particles must be small enough:
\begin{equation}
r<(\mu_B^2/kT_c)^{1/3}\approx a_B,\label{L1}
\end{equation}
where $a_B=\frac{\hbar^2}{m_e e^2}$ is the Bohr radius.

That is,  two collectivized electrons must be localized in one lattice site volume.
It is  agreed that the superconductivity can occur only in metals with two collectivized
electrons per atom, and cannot exist in the monovalent alkali and noble metals.

It is easy to see that the presence of  magnetic moments on  ion sites  should interfere with the magnetic combination of electrons. This is confirmed by the experimental fact: as there are no  strong magnetic substances among superconductors, so adding of iron, for example, to traditional superconducting alloys always leads to a lower  critical temperature.

On the other hand, this magnetic coupling should not be destroyed at the critical temperature. The energy of interaction between two electrons, located near one lattice site, can be much greater.
This is confirmed by experiments  showing that throughout the period of the magnetic flux quantization, there is no change at the transition through the critical temperature of superconductor \cite{Shab}, \cite{Sharv}.

The outcomes of these experiments are evidence that the existence of the mechanism of electron pairing is a necessary but not a sufficient condition for the existence of superconductivity.

The magnetic mechanism of electronic pairing proposed above  can be seen as an assumption which is consistent with the measurement data and therefore needs a more detailed theoretic consideration and further refinement.

 On the other hand, this issue is not very important in the grander scheme, because the nature of the mechanism that causes electron pairing is not of a significant importance.
 Instead, it is important that there is a mechanism which converts the electronic gas into an ensemble of charged bosons with zero spin in the considered temperature range (as well as in a some range of temperatures above $T_c$).

 If the temperature is not low enough, the electronic pairs still exist but their zero-point oscillations  are disordered. Upon reaching the $T_c$, the interaction between  zero-point oscillations should cause their ordering  and therefore a  superconducting state is created.

\section{The interaction of zero-point oscillations}
The principal condition for the superconducting state formation is the ordering of zero-point oscillations. It is realized
because the paired electrons obeying Bose-Einstein statistics attract each other.

The origin of this attraction can be explained as follows.

\begin{figure}
\hspace{1.5cm}
\includegraphics[scale=0.5]{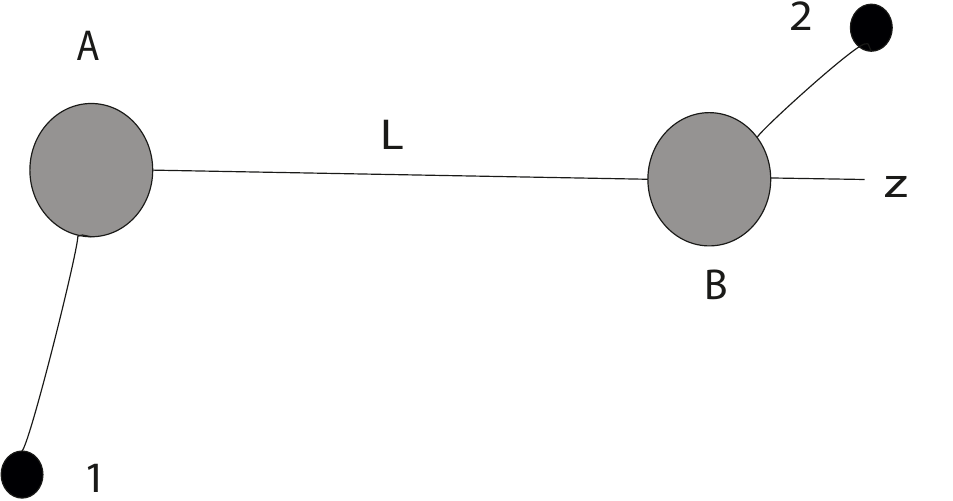}
\caption{Two ions placed on the distance $L$ and centers of their electronic clouds.}\label{v2}
\end{figure}

Let two  ion A and B be located on the z axis at the distance L from each other.
Two  collectivized electrons create clouds with centers at points 1 and 2 in the vicinity of each ions (Figure{\ref{v2}}).
Let $r_1$ be the radius-vector of the center of the first electronic cloud relative to the ion A and $ r_2 $ is the radius-vector of the second electron relative to the ion B.

Following the Born-Oppenheimer approximation, slowly oscillating ions are assumed fixed.
Let the temperature be low enough $ (T \rightarrow 0) $, so  only zero-point fluctuations of electrons would be taken into consideration.

In this case, the Hamiltonian of the system can be written as:
\begin{equation}
\begin{array}{l}
H=H_0+H'
\\
\\
H_0=-\frac{\hbar^2}{4m_e}\left(\nabla_1^2+\nabla_2^2\right)-
\\
\qquad \qquad \qquad \qquad {}
-\frac{4e^2}{r_1}-\frac{4e^2}{r_2}\\
\\
H'=\frac{4e^2}{L}+\frac{4e^2}{r_{12}}
{-\frac{4e^2}{r_{1B}}-\frac{4e^2}{r_{2A}}}
\label{LL1}
\end{array}
\end{equation}
Eigenfunctions of the unperturbed Hamiltonian describes two ions surrounded by electronic clouds without interactions between them.
Due to the fact that the distance between the ions  is large compared with the size of the electron clouds $L\gg r$ , the additional term $H'$ characterizing the interaction can be regarded as a perturbation.

If we are interested in  the leading term of the interaction energy for L, the function  $ H '$ can be expanded in a series in powers of $1/L$ and we can write the first term:
\begin{equation}
\begin{array}{l}
H'=\frac{4e^2}{L} \bigg\{1 +
\bigg[
1+\frac{2(z_2-z_1)}{L}+
\\
+\frac{(x_2-x_1)^2+(y_2-y_1)^2+(z_2-z_1)^2}{L^2}
\bigg]^{-1/2}- \\
\qquad \qquad
-\left(1-\frac{2z_1}{L}+\frac{r_1^2}{L^2}\right)^{-1/2}-\\
\qquad \qquad
-\left(1+\frac{2z_2}{L}+\frac{r_2^2}{L^2}\right)^{-1/2}
\bigg\}.
\end{array}
\end{equation}
After combining the terms in this expression, we get:
\begin{equation}
H'\approx\frac{4e^2}{L^3}\left(x_1x_2+y_1y_2-2z_1z_2\right).\label{h1}
\end{equation}
This expression describes the interaction of two dipoles $ d_1 $ and $ d_2 $, which are formed by  fixed ions and electronic clouds of the corresponding instantaneous configuration.

Let us determine the displacements of  electrons which lead to an attraction in the system .

Let  zero-point fluctuations of the dipole moments formed by ions with their electronic clouds  occur with the  frequency $\Omega_0$, whereas each dipole moment can be decomposed into three orthogonal projection
$d_x=ex, d_y=ey $ and $d_z=ez$, and fluctuations of the second clouds are shifted in phase on $\varphi_x, \varphi_y $ and $\varphi_z$ relative to
fluctuations of the first.

As can be seen from Eq.(\ref{h1}), the interaction of z-components  is advantageous at in-phase  zero-point oscillations of  clouds,  i.e., when $\varphi_z = 2 \pi$.

Since the interaction of oscillating electric dipoles is due to the occurrence of oscillating electric field generated by them, the phase shift on $2\pi$ means that attracting  dipoles are placed  along the z-axis on the  wavelength $\Lambda_0$:
\begin{equation}
L_z=\Lambda_0=\frac{c}{2\pi \Omega_0}.\label{Lz1}
\end{equation}
As follows from (\ref{h1}), the attraction of dipoles at the interaction of the x and y-component will occur if these oscillations are in antiphase, i.e. if the dipoles are separated along these axes on the distance equals to half of the wavelength:
\begin{equation}
L_{x,y}=\frac{\Lambda_0}{2}=\frac{c}{4\pi \Omega_0}.\label{Lxy}
\end{equation}
In this case
\begin{equation}
H'=-{4e^2}\left(\frac{x_1x_2}{L_x^3}+\frac{y_1y_2}{L_y^3}+2\frac{z_1z_2}{L_z^3}\right).\label{Lz3}
\end{equation}
Assuming that the electronic clouds have isotropic oscillations with amplitude $ a_0 $ for each axis
\begin{equation}
x_1=x_2=y_1=y_2=z_1=z_2=a_0
\end{equation}
we obtain
\begin{equation}
H'=576\pi^3\frac{e^2}{c^3}\Omega_0^3 a_0^2.\label{hh}
\end{equation}

\section{The zero-point oscillations amplitude}
The principal condition for the superconducting state formation, that is the ordering of zero-point oscillations, is realized
due to the fact that the paired electrons, which obey Bose-Einstein statistics, interact  with each other.

At they interact, their amplitudes, frequencies and phases of zero-point oscillations become ordered.

Let an electron gas has density $n_e$ and its Fermi-energy be $\mathcal{E}_F$. Each electron of this gas can be considered as fixed inside a cell with linear dimension $\lambda_F$:\footnote{Of course, the electrons are quantum particles and their fixation cannot be considered too literally. Due to the Coulomb forces of ions, it is more favorable for  collectivized electrons  to be placed near the ions  for the shielding  of ions fields. At the same time, collectivized electrons are spread over whole metal.
It is wrong to think that a particular electron is fixed inside a cell near to a particular ion.
But the spread of the electrons does not play a fundamental importance for our further consideration, since  there are two electrons near the node of the lattice in the divalent metal at any given time.
They  can be considered as located inside the cell as averaged.}
\begin{equation}
\lambda_F^3=\frac{1}{n_e}
\end{equation}
which corresponds to the de Broglie wavelength:
\begin{equation}
\lambda_F=\frac{2\pi \hbar}{p_F}.\label{lF1}
\end{equation}
Having taken into account (\ref{lF1}), the Fermi energy of the electron gas can be written as
\begin{equation}
\mathcal{E}_F=\frac{p_F^2}{2m_e}=2\pi^2\frac{e^2a_B}{\lambda_F^2}.\label{EF2}
\end{equation}

However, a free electron interacts with the ion at its zero-point oscillations.
If we consider the  ions system as a positive background uniformly spread over the cells, the electron inside one cell  has the potential energy:
\begin{equation}
\mathcal{E}_p\simeq -\frac{e^2}{\lambda_F}.
\end{equation}
As zero-point oscillations of the electron pair are   quantized by definition, their frequency and amplitude are related
\begin{equation}
{m_e a_0^2\Omega_0}\simeq \frac{\hbar}{2}.
\end{equation}
Therefore, the kinetic energy of electron undergoing zero-point oscillations in a limited region of space, can be written as:
\begin{equation}
\mathcal{E}_k\simeq \frac{\hbar^2}{2m_e a_0^2}.
\end{equation}

In accordance with the virial theorem \cite{vir}, if a particle executes a finite motion,
its potential energy $\mathcal{E}_p$ should be associated with its kinetic energy $\mathcal{E}_k$ through the simple relation $|\mathcal{E}_p|=2\mathcal{E}_k$.

In this regard, we find that the amplitude of the zero-point oscillations of an electron in a cell is:
\begin{equation}
a_0\simeq \sqrt{2\lambda_F a_B}. \label{a0}
\end{equation}

\section{The condensation temperature}
Hence the interaction energy, which unites  particles into the condensate of ordered zero-point oscillations
\begin{equation}
\Delta_0\equiv H'=18\pi^3\alpha^3\frac{e^2a_B}{\lambda_F^2},\label{Dh1}
\end{equation}
where $\alpha=\frac{1}{137}$ is the fine structure constant.

Comparing this  association energy   with the Fermi energy (\ref{EF2}), we obtain
\begin{equation}
\frac{\Delta_0}{\mathcal{E}_F}= 9\pi \alpha^3\simeq 1.1\cdot 10^{-5}.\label{Dh2}
\end{equation}

Assuming that the critical temperature below which the possible existence of such condensate is approximately equal
\begin{equation}
T_c\simeq \frac{1}{2}\frac{\Delta_0}{k}
\end{equation}
(the coefficient approximately equal to 1/2 corresponds to the experimental data, discussed below in the section (\ref{Delta-TcE})).

After substituting obtained parameters, we have
\begin{eqnarray}
T_c\simeq 5.5\cdot 10^{-6}{T}_F\label{TT0}
\end{eqnarray}

The experimentally measured ratios  $\frac{T_c}{T_F}$  for I-type superconductors are given in Table (\ref{de1}) and in Fig.(\ref{TT0g}).

The straight line on this figure is obtained from Eq.(\ref{TT0}), which
as seen defines an upper limit of critical temperatures of I-type superconductors.

\vspace{0.5cm}

\newpage

\chapter[The condensate and type-I superconductors]{The condensate of zero-point oscillations  and type-I superconductors}
\section{The critical temperature of type-I superconductors}
In order to compare the critical temperature of the condensate of zero-point oscillations with   measured critical temperatures of superconductors, at first we should make an estimation on the Fermi energies of superconductors.
For this we use the experimental data for the Sommerfeld`s constant  through which the Fermi energy can be expressed:
\begin{equation}
\gamma=\frac{\pi^2  k^2 n_e}{4\mathcal{E}_F}=\frac{1}{2}\cdot\left(\frac{\pi}{3}\right)^{2/3}\left(\frac{k}{\hbar}\right)^2 m_e n_e^{1/3}\label{gz}
\end{equation}
So on the basis of Eqs.(\ref{EF2}) and (\ref{gz}), we get:
\begin{equation}
kT_F(\gamma)=\frac{p_F^2(\gamma)}{2m_e}\simeq \left(\frac{12}{k^2}\right)^2\left(\frac{\hbar^2}{2m_e}\right)^3\gamma^2.\label{kTF}
\end{equation}

On base of these calculations we obtain possibility to relate  directly the critical temperature of a superconductor with the experimentally measurable parameter: with its electronic specific heat.

Taking into account Eq.(\ref{TT0}), we have:
\begin{equation}
\Delta_0\simeq \Theta\gamma^2\label{tcc},
\end{equation}
where the constant
\begin{equation}
\Theta\simeq 31\frac{\pi^2}{k} \left[\frac{\alpha\hbar^2}{k m_e}\right]^3\simeq 6.65\cdot 10^{-22}\frac{K^4 cm^6}{erg}\label{tc2}.
\end{equation}

\begin{figure}
\hspace{0.5cm}
\includegraphics[scale=0.5]{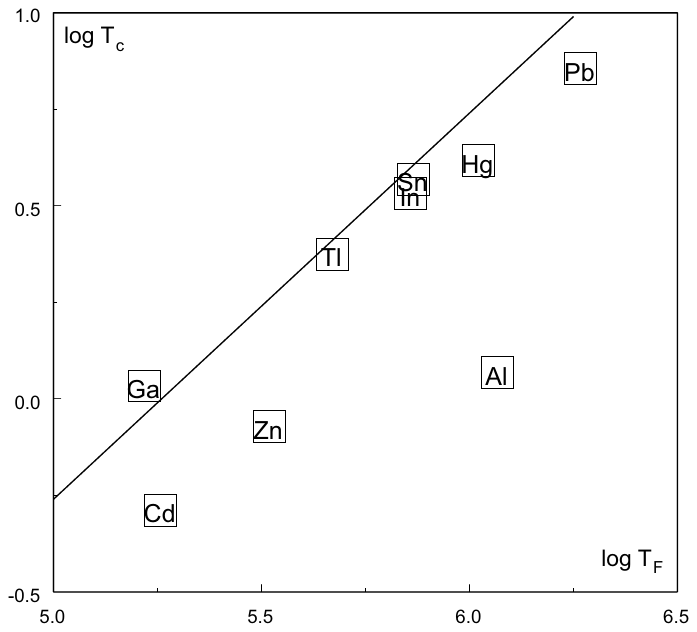}
\vspace{-7cm}\caption {The comparison of critical temperatures $T_c$ of type-I superconductors with their Fermi temperatures $T_F$. The straight line is obtained from Eq.(\ref{TT0}).}\label{TT0g}
\end{figure}

The comparison of the calculated parameters and  measured data ({\cite{Ketterson}},{\cite{Pool}})  is given in Table (\ref{de1})-(\ref{2tt}) and in Fig.({\ref{TT0g}}) and (\ref{tc2g}).\\
\bigskip

\begin{table}
\centering
\begin{tabular}{||c|c|c|c||}\hline\hline
&&&\\
  superconductor &$T_c$,K&$T_F$,K&$\frac{T_c}{T_F}$\\
  &&Eq(\ref{kTF})&\\
    &&&\\\hline
  Cd &0.51&$1.81\cdot 10^5$&$2.86\cdot 10^{-6}$\\
  Zn &0.85&$3.30\cdot 10^5$&$2.58\cdot 10^{-6}$\\
  Ga &1.09&$1.65\cdot 10^5$&$6.65\cdot 10^{-6}$\\
  Tl &2.39&$4.67\cdot 10^5$&$5.09\cdot 10^{-6}$\\
  In &3.41&$7.22\cdot 10^5$&$4.72\cdot 10^{-6}$\\
  Sn &3.72&$7.33\cdot 10^5$&$5.08\cdot 10^{-6}$\\
  Hg &4.15&$1.05\cdot 10^6$&$3.96\cdot 10^{-6}$\\
  Pb &7.19&$1.85\cdot 10^6$&$3.90\cdot 10^{-6}$\\ \hline\hline
  \end{tabular}
  \caption{The comparison of the calculated values of  superconductors critical temperatures with measured Fermi temperatures}
\label{de1}
\end{table}

\bigskip

\begin{table}
\centering
\begin{tabular}{||c|c|c|c|c||}\hline\hline
  super-&$T_c$(measur),&$\gamma,\frac{erg}{cm^3K^2}$&$T_c$(calc),K&$\frac{T_c (calc)}{T_c(meas)}$\\
conductors  &K&&Eq.(\ref{tcc})&\\\hline
  Cd &$0.517$&532&$0.77$&1.49\\
  Zn &$0.85$&718&$1.41$&1.65\\
  Ga &$1.09$&508&$0.70$&0.65\\
  Tl &$2.39$&855&$1.99$&0.84\\
  In &$3.41$&1062&$3.08$&0.90\\
  Sn &$3.72$&1070&$3.12$&0.84\\
  Hg &$4.15$&1280&$4.48$&1.07\\
  Pb &$7.19$&1699&$7.88$&1.09\\ \hline\hline
\end{tabular}
\caption{The comparison of the calculated values of superconductors critical temperatures with measurement data}
\label{2tt}
\end{table}

\bigskip

\bigskip

\bigskip

\section[The relation of critical parameters]{The relation of critical parameters of type-I superconductors}
\label{crit-param1}
The phenomenon of condensation of zero-point oscillations in the electron gas has its characteristic features.

 There are several ways of destroying  the zero-point oscillations condensate in electron gas:

 Firstly, it can be evaporated by heating. In this case, evaporation of the condensate should possess the properties of an order-disorder transition.

 Secondly, due to the fact that the oscillating electrons carry electric charge, the condensate can be destroyed by the application of a sufficiently strong magnetic field.

 For this reason, the critical temperature and critical magnetic field of the condensate will be interconnected.

  This interconnection should manifest itself through the relationship of the critical temperature and critical field of the superconductors, if superconductivity occurs as result of an ordering of zero-point fluctuations.

Let us assume that at a given temperature ${T <T_{c}}$ the system of vibrational levels of conducting electrons consists of only two levels:

 firstly, basic level which is characterized by an anti-phase oscillations of the electron pairs at the distance $\Lambda_0/2$, and

 secondly, an excited level characterized by in-phase oscillation of the pairs.

Let the population of the basic level be $N_0$ particles and the excited level has  $N_1$ particles.

Two electron pairs at an in-phase oscillations have a high energy of interaction and therefore cannot form the condensate.
The condensate can be formed only by the particles that make up the difference between the populations of levels $N_0-N_1$.
In a dimensionless form, this difference defines the order parameter:
\begin{equation}
\Psi=\frac{N_0}{N_0+N_1}-\frac{N_1}{N_0+N_1}.
\end{equation}
In the theory of superconductivity, by definition, the order parameter is determined by the value of the energy gap
\begin{equation}
\Psi=\Delta_T/\Delta_0.
\end{equation}
When taking a counting of energy from the level $\varepsilon_0$, we obtain
\begin{equation}
\frac{\Delta_{T}}{\Delta_0}=\frac{N_0-N_1}{N_0+N_1}\simeq\frac{e^{2\Delta_{T}/kT} -1}{e^{2\Delta_{T}/kT} +1}=th(2\Delta_{T}/kT).\label{det}
\end{equation}
Passing to dimensionless variables $\delta\equiv \frac{\Delta_{T}}{\Delta_0}$ , $t\equiv \frac{kT}{kT_c}$
and $\beta\equiv\frac{2\Delta_0}{kT_c}$ we have
\begin{equation}
\delta=\frac{e^{\beta\delta/t} -1}{e^{\beta\delta/t} +1}=th(\beta\delta/t).\label{del}
\end{equation}
This equation describes the temperature dependence of the energy gap in the spectrum of zero-point oscillations.
It is similar to other equations describing other physical phenomena, that are also characterized by the existence of the temperature dependence of order parameters \cite{LL},\cite{Kit}. For example, this dependence is similar to temperature dependencies of  the concentration of the superfluid component in liquid helium or the spontaneous magnetization of ferromagnetic materials. This equation is the same for all order-disorder transitions (the phase transitions of 2nd-type in the Landau classification).

\begin{figure}
\centering
\includegraphics[scale=0.75]{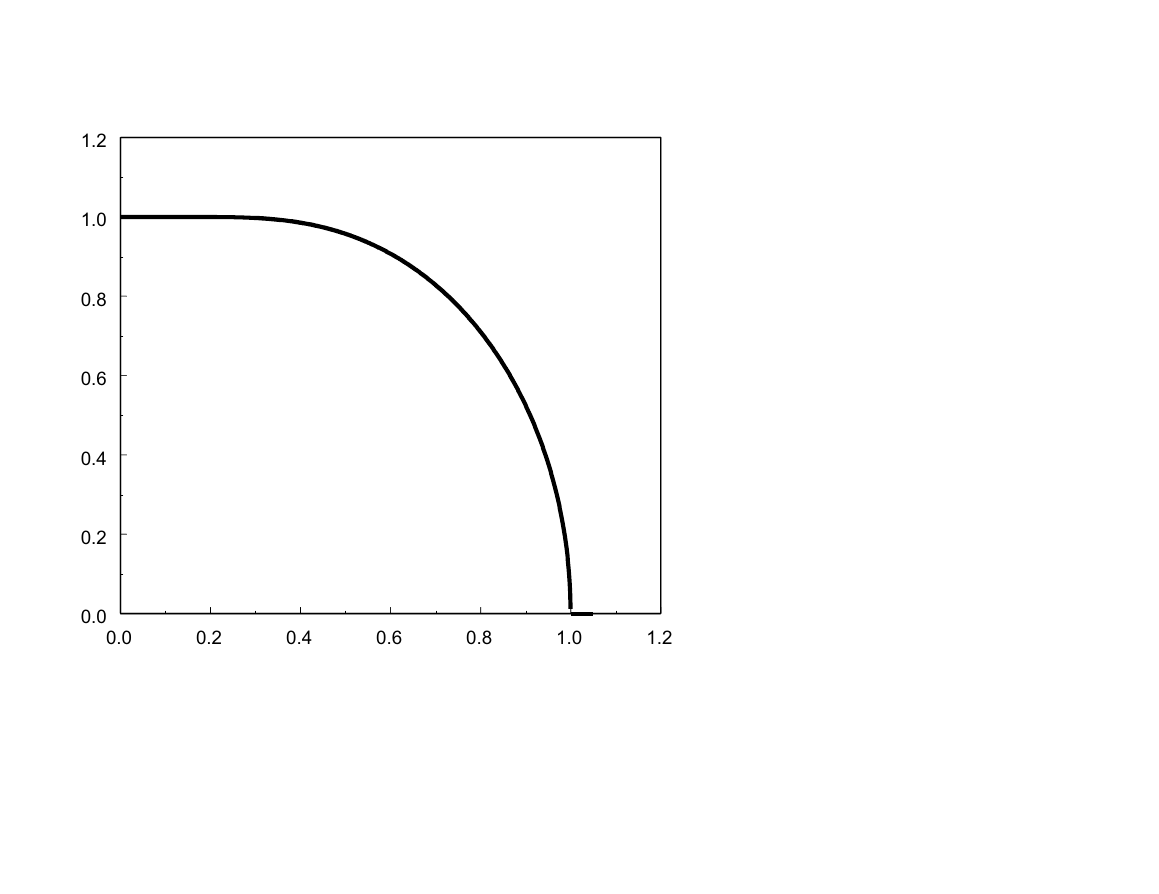}
\vspace{-3cm}\caption{The temperature dependence of the value of the gap in the energetic spectrum of zero-point oscillations calculated on Eq.(\ref{del}).}
\label{D-T}
\end{figure}

The solution of this equation, obtained by the iteration method, is shown in Fig.(\ref{D-T}).

This decision is in a agreement with  the known  transcendental equation of the BCS, which was obtained by the integration of the phonon spectrum, and is in a satisfactory agreement with the measurement data.

After numerical integrating we can obtain the averaging value of the gap:
\begin{equation}
\langle\Delta\rangle=\Delta_0\int_0^1 \delta dt=0.852~\Delta_0~.\label{0.8}
\end{equation}

To convert the condensate into the normal state, we  must  raise half of its particles  into the excited state  (according to Eq.(\ref{det}), the gap collapses under this condition). To do this, taking into account Eq.(\ref{0.8}), the unit volume of condensate   should have the energy:
\begin{equation}
\mathcal{E}_T\simeq  \frac{1}{2} n_0 \langle\Delta_0  \rangle  \approx  \frac{0.85}{2}\left(\frac{m_e}{2\pi^2\alpha\hbar^2}\right)^{3/2}\Delta_0^{5/2},\label{ET}
\end{equation}
On the other hand, we can obtain the normal state of an electrically charged condensate when applying a magnetic field of critical value $H_c$ with the density of energy:
\begin{equation}
\mathcal{E}_H= \frac{H_c^2}{8\pi}.\label{EH}
\end{equation}
As a result, we acquire the condition:
\begin{equation}
\frac{1}{2}n_0 \langle\Delta_0  \rangle=\frac{H_c^2}{8\pi}.\label{TH1}
\end{equation}
This creates a  relation of  the critical temperature to the critical magnetic field of the zero-point oscillations condensate of the charged bosons.

\begin{figure}
\includegraphics[scale=0.5]{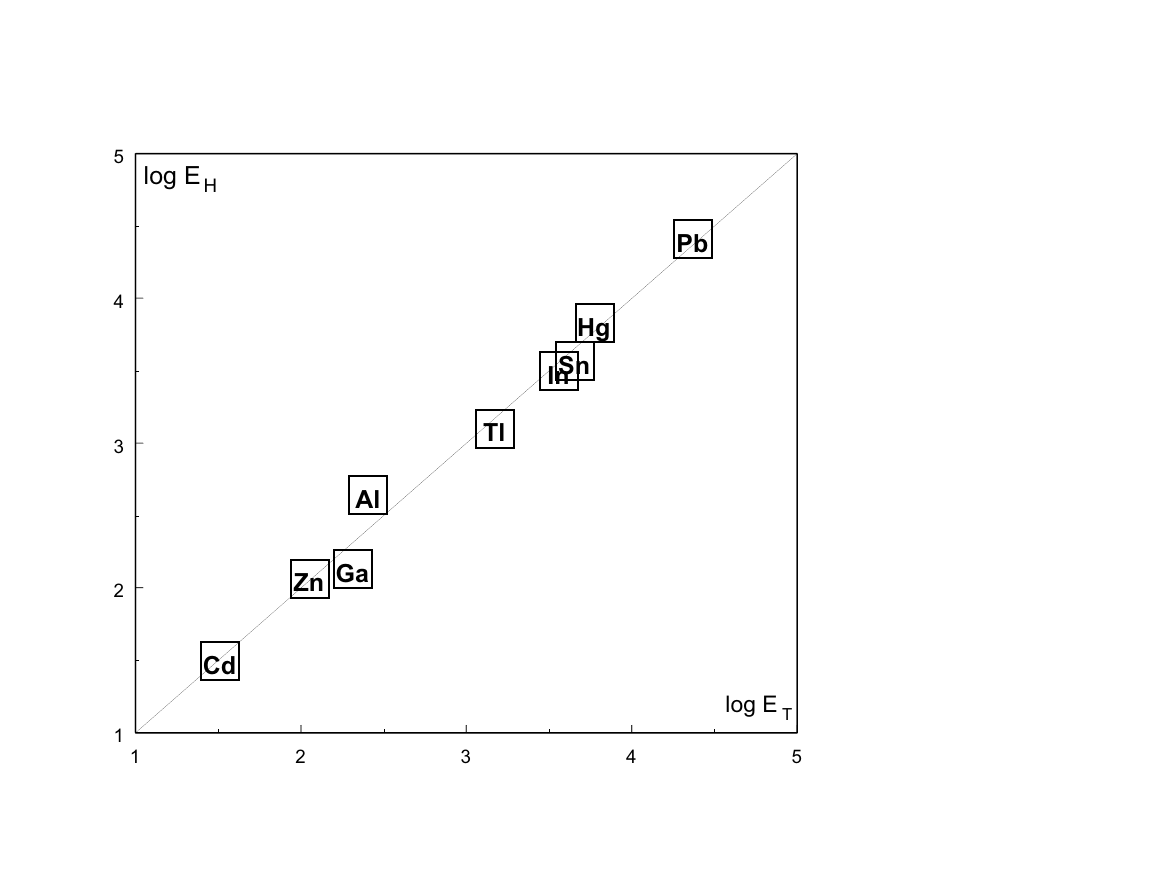}
\caption{The comparison of the critical energy densities $\mathcal{E}_T$ (Eq.(\ref{ET})) and $\mathcal{E}_H$ (Eq.(\ref{EH})) for the type-I superconductors.}\label{eh-et2}
\end{figure}

The comparison of the critical energy densities $\mathcal{E}_T$ and $\mathcal{E}_H$ for type-I superconductors are shown in Fig.(\ref{eh-et2}).
As shown,  the obtained agreement between  the energies $\mathcal{E_T}$ (Eq.(\ref{ET})) and $\mathcal{E_H}$ (Eq.(\ref{EH}))  is  quite satisfactory for type-I superconductors \cite{Pool},\cite{Ketterson}.
 A similar comparison for type-II superconductors shows results that differ by a factor two approximately.
 The reason for this will be considered below.
 The correction of this calculation, has not apparently made sense here.
 The purpose of these calculations was to show that the description of superconductivity as the effect of the condensation of ordered zero-point oscillations is in accordance with the available experimental data. This goal is considered reached in  the simple case of type-I superconductors.

\section[The critical magnetic field]{The critical magnetic field of superconductors}
The direct influence of the external magnetic field of the critical value applied to the electron system is too weak to disrupt the dipole-dipole interaction of two paired electrons:
\begin{equation}
\mu_B H_c \ll kT_c.\label{L2}
\end{equation}
In order to violate the superconductivity,   the ordering of the electron zero-point oscillations must be  destroyed.
 For this the presence of relatively weak magnetic field is required.

At combing of Eqs.(\ref{TH1}),(\ref{ET}) and (\ref{a0}), we can express the gap through the critical magnetic field and the magnitude of the oscillating dipole moment:
\begin{equation}
\Delta_0\approx \frac{1}{2}~e~a_0~H_c.\label{dah}
\end{equation}
The properties of the zero-point oscillations of the electrons should not be dependent on the characteristics of the mechanism of association and also on the condition of the existence of electron pairs. Therefore, we should expect that this equation would also be valid  for type-I superconductors, as well as for II-type superconductors (for II-type superconductor  $H_c=H_{c1}$ is the first critical field)

\begin{figure}
\vspace{-13.5cm}
\centering
\includegraphics[scale=0.8]{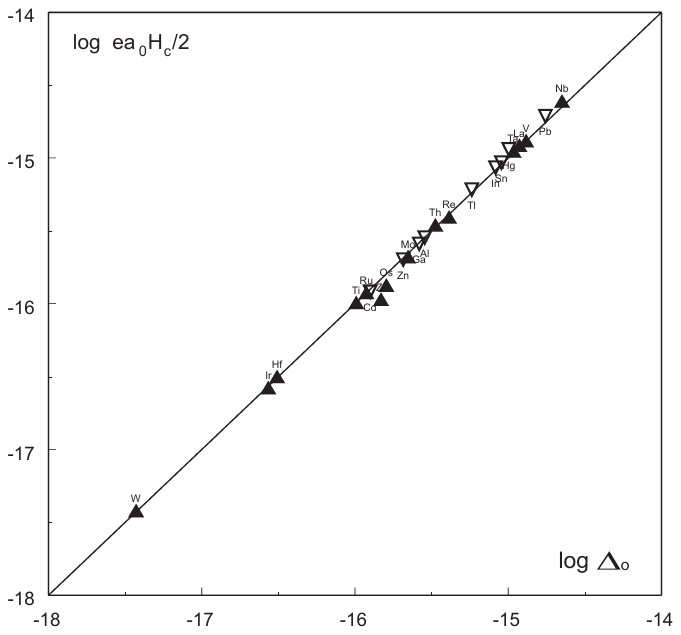}
\vspace{-2cm}\caption {The comparison of the calculated energy of superconducting pairs in the critical magnetic field with the value of the superconducting gap.
Here, the following key applies:
filled triangles - type-II superconductors,
empty triangles - type-I superconductors.
On vertical axis - logarithm of the product of the calculated value of the oscillating  dipole moment of an electron pair on the critical magnetic field is plotted.
On horizontal axis - the value of the gap is shown.}\label{rd2}
\end{figure}

An agreement with this condition is illustrated on the Fig.(\ref{rd2}).

\bigskip

\section{The density of superconducting carriers}
Let us consider the process of heating the electron gas in metal.
 When heating, the electrons from levels slightly below the Fermi-energy are raised to higher levels. As a result, the levels closest to the Fermi level, from which at low temperature electrons were forming bosons, become vacant.

At critical temperature $T_c$, all electrons from the levels of energy bands from $\mathcal{E}_F-\Delta$ to  $\mathcal{E}_F$  move to higher levels (and the gap collapses). At this temperature superconductivity is therefore destroyed completely.

This band of energy can be filled by $N_\Delta$ particles:
\begin{equation}
N_\Delta=2\int_{\mathcal{E}_F-\Delta}^{\mathcal{E}_F}
F(\mathcal{E})D(\mathcal{E})d\mathcal{E}\label{ne}.
\end{equation}
Where  $F(\mathcal{E})=\frac{1}{e^{\frac{\mathcal{E}-\mu}{\tau}}+1}$ is the Fermi-Dirac function and $D(\mathcal{E})$ is  number of states per an unit energy interval, a deuce front of the integral arises from the fact that there are two electron at each energy level.

To find the density of states $D(\mathcal{E})$,  one needs to find the difference in energy of the system at $T=0$ and finite temperature:
\begin{equation}
\Delta \mathcal{E} =\int_0^\infty F(\mathcal{E})\mathcal{E}D(\mathcal{E})d\mathcal{E}-\int_0^{\mathcal{E}_F} \mathcal{E}D(\mathcal{E})d\mathcal{E}\label{dE}.
\end{equation}
 For the calculation of the density of states $D(\mathcal{E})$, we must note that two electrons can be placed on each level. Thus, from the expression of the Fermi-energy Eq.(\ref{EF2})
we obtain
\begin{equation}
D({E}_F)=\frac{1}{2}\cdot\frac{dn_e}{d\mathcal{E}_F}=\frac{3n_e}{4\mathcal{E}_F}=\frac{3\gamma}{2k^2\pi^2},\label{d}
\end{equation}
where
\begin{equation}
\gamma=\frac{\pi^2  k^2 n_e}{4\mathcal{E}_F}=\frac{1}{2}\cdot\left(\frac{\pi}{3}\right)^{3/2}\left(\frac{k}{\hbar}\right)^2 m_e n_e^{1/3}\label{gzz}
\end{equation}
is the Sommerfeld constant
\footnote{It should be noted that because on each level two electrons can be placed, the expression for the Sommerfeld constant Eq.(\ref{gzz}) contains the additional factor $1/2$ in comparison with the usual formula in  literature \cite{Kit}}.

Using similar arguments, we can calculate the number of electrons, which populate the levels in the range from $\mathcal{E}_F-\Delta$ to $\mathcal{E}_F$. For an unit volume of material, Eq.(\ref{ne}) can be rewritten as:
\begin{equation}
n_\Delta=2kT\cdot D(\mathcal{E}_F)\int_{-\frac{\Delta_0}{kT_c}}^0  \frac{dx}{(e^x +1)}. 
\end{equation}

By supposing that for superconductors $\frac{\Delta_0}{kT_c}=1.86$, as a result of numerical integration we obtain
\begin{equation}
\int_{-\frac{\Delta_0}{kT_c}}^0  \frac{dx}{(e^x +1)}=\left[x-ln(e^x+1)\right]_{-1.86}^0\approx 1.22 . 
\end{equation}
Thus, the density of electrons, which throw up above the Fermi level in a metal at temperature $T = T_c$ is
\begin{equation}
n_e(T_c)\approx 2.44 \left(\frac{3\gamma}{k^2\pi^2}\right)kT_c.\label{net}
\end{equation}
Where the Sommerfeld constant $\gamma$ is  related to the volume unit  of the metal.

From Eq.(\ref{Lxy}) it follows
\begin{equation}
L_0\simeq\frac{\lambda_F}{\pi\alpha}\label{L0}
\end{equation}
 and this  forms the ratio of the condensate particle density  to the Fermi gas density:
\begin{equation}
\frac{n_0}{n_e}=\frac{\lambda_F^3}{L_0^3}\simeq\left(\pi\alpha\right)^3\simeq 10^{-5}.\label{n0-ne}
\end{equation}
When using these equations, we can find a linear dimension of localization for an electron pair:
\begin{equation}
L_0 =\frac{\Lambda_0}{2}\simeq  \frac{1}{\pi\alpha(n_e)^{1/3}}.\label{L}
\end{equation}
or, taking into account Eq.(\ref{a0}), we can obtain the relation between the density of particles in the condensate and the value of the energy gap:
\begin{equation}
\Delta_0\simeq 2\pi^2\alpha\frac{\hbar^2}{m_e}n_0^{2/3}\label{D-2}
\end{equation}
or
\begin{equation}
n_0=\frac{1}{L_0^3}=\left(\frac{m_e}{2\pi^2\alpha \hbar^2}\Delta_0\right)^{3/2}.\label{n-0}
\end{equation}
It should be noted that the obtained ratios for the zero-point oscillations condensate (of bose-particles)
differ from the corresponding expressions for the bose-condensate of particles, which can be obtained in many courses (see eg \cite{LL}). The expressions for the ordered condensate of zero-point  oscillations have an additional coefficient $\alpha$ on the right side of Eq.(\ref{D-2}).

\vspace{0.5cm}
The de Broglie wavelengths of  Fermi electrons expressed through the Sommerfelds constant
\begin{equation}
\lambda_F=\frac{2\pi \hbar}{p_F(\gamma)}\simeq\frac{\pi}{3}\cdot\frac{k^2 m_e}{\hbar^2 \gamma}\label{lF}
\end{equation}
are shown in Tab.\ref{n0e}.

In accordance with Eq.(\ref{L0}), which was obtained at the zero-point oscillations consideration,  the ratio $\frac{\lambda_F}{\Lambda_0}\simeq 2.3\cdot 10^{-2}$.

In connection with this ratio, the calculated ratio of the zero-point oscillations condensate density to the density of fermions in accordance with Eq.(\ref{n0-ne}) should be near to $10^{-5}$.

 It can be therefore be seen, that calculated estimations of the condensate parameters  are in satisfactory agreement with experimental data of superconductors.
{
\begin{table}
\centering
\begin{tabular}{||c|c|c|c|c||}\hline\hline
&&&&\\
  super-&$\lambda_F$,cm&$\Lambda_0$,cm&$\frac{\lambda_F}{\Lambda_0}$&
  $\frac{n_0}{n_e}=\left(\frac{\lambda_F}{\Lambda_0}\right)^3$\\
   conductor &Eq(\ref{lF})&Eq(\ref{Lxy})&&\\
   &&&&\\\hline
  Cd &$3.1\cdot 10^{-8}$&$1.18\cdot 10^{-6}$&$2.6\cdot 10^{-2}$&$1.8\cdot 10^{-5}$\\
  Zn &$2.3\cdot 10^{-8}$&$0.92\cdot 10^{-6}$&$2.5\cdot 10^{-2}$&$1.5\cdot 10^{-5}$\\
  Ga &$3.2\cdot 10^{-8}$&$0.81\cdot 10^{-6}$&$3.9\cdot 10^{-2}$&$6.3\cdot 10^{-5}$\\
  Tl &$1.9\cdot 10^{-8}$&$0.55\cdot 10^{-6}$&$3.4\cdot 10^{-2}$&$4.3\cdot 10^{-5}$\\
  In &$1.5\cdot 10^{-8}$&$0.46\cdot 10^{-6}$&$3.2\cdot 10^{-2}$&$3.8\cdot 10^{-5}$\\
  Sn &$1.5\cdot 10^{-8}$&$0.44\cdot 10^{-6}$&$3.4\cdot 10^{-2}$&$4.3\cdot 10^{-5}$\\
  Hg &$1.3\cdot 10^{-8}$&$0.42\cdot 10^{-6}$&$3.1\cdot 10^{-2}$&$2.9\cdot 10^{-5}$\\
  Pb &$1.0\cdot 10^{-8}$&$0.32\cdot 10^{-6}$&$3.1\cdot 10^{-2}$&$2.9\cdot 10^{-5}$\\ \hline\hline
  \end{tabular}
 \caption{The ratios $\frac{\lambda_F}{\Lambda_0}$ and $\frac{n_0}{n_e}$ for type-I superconductors}
\label{n0e}
\end{table}


\vspace{0.5cm}

Based on these calculations, it is interesting to compare the density of superconducting carriers
$n_0$  at $T = 0$, which is described by Eq.(\ref{n-0}), with the density of normal carriers $n_e(T_c)$, which are
evaporated on levels above $\mathcal{E}_F$ at $T=T_c$ and are described by Eq.(\ref{net}).
\bigskip

\begin{figure}
\hspace{1.5cm}
\includegraphics[scale=0.5]{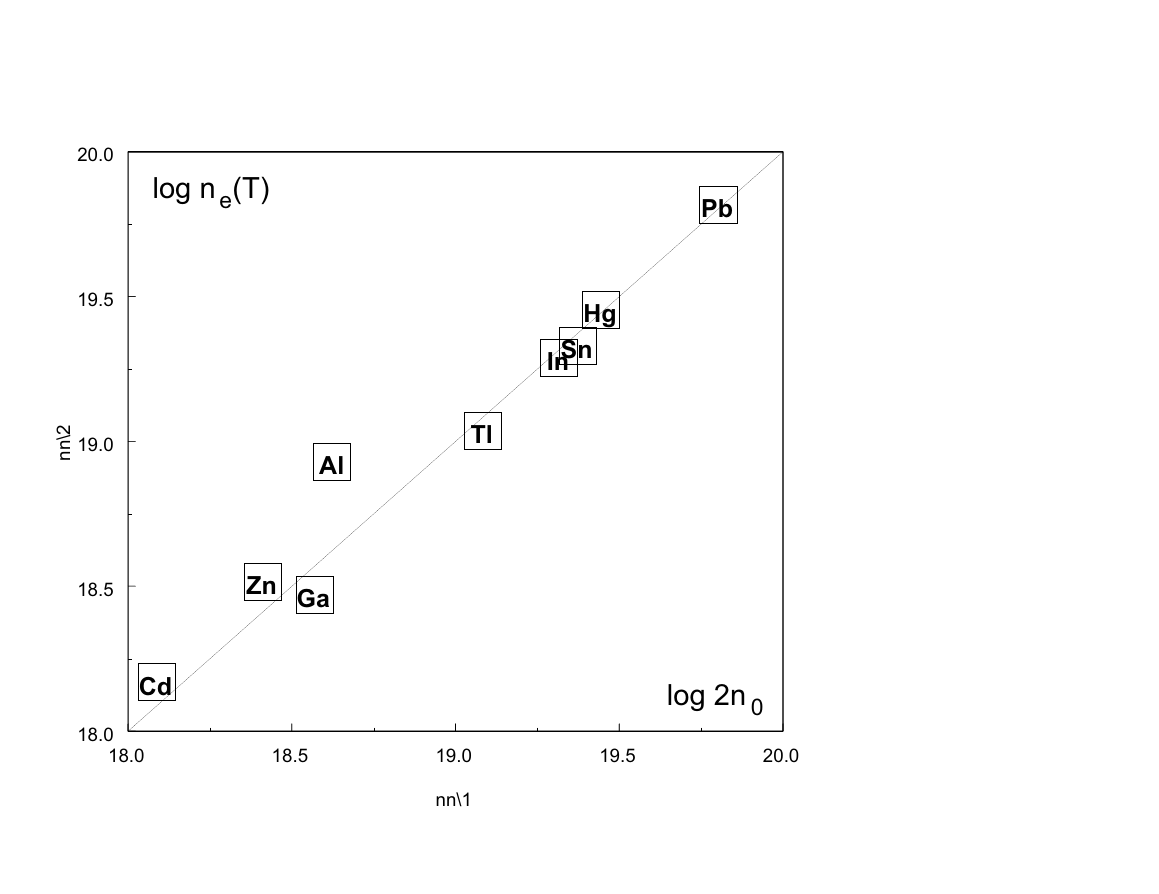}
\caption {The comparison of the number of superconducting carriers at $T=0$ with the number of thermally activated electrons at $T=T_c$.}\label{n0Ne}
\end{figure}

\bigskip

This comparison is shown in Table ({\ref{nna}}) and Fig.{\ref{n0Ne}}.
 (Data has been taken from the tables {\cite{Pool}}, {\cite{Ketterson}}).

\bigskip
\begin{table}
\centering
\begin{tabular}{||c|c|c|c||}\hline\hline
  superconductor&$n_0$&$n_e({T_c})$&$2n_0/n_e(T_c)$\\\hline
  Cd &$6.11\cdot 10^{17}$&$1.48\cdot 10^{18}$&0.83\\
  Zn &$1.29\cdot 10^{18}$&$3.28\cdot 10^{18}$&0.78\\
  Ga &$1.85\cdot 10^{18}$&$2.96\cdot 10^{18}$&1.25\\
  Al &$2.09\cdot 10^{18}$&$8.53\cdot 10^{18}$&0.49\\
  Tl &$6.03\cdot 10^{18}$&$1.09\cdot 10^{19}$&1.10\\
  In &$1.03\cdot 10^{19}$&$1.94\cdot 10^{19}$&1.06\\
  Sn &$1.18\cdot 10^{19}$&$2.14\cdot 10^{19}$&1.10\\
  Hg &$1.39\cdot 10^{19}$&$2.86\cdot 10^{19}$&0.97\\
  Pb &$3.17\cdot 10^{19}$&$6.58\cdot 10^{19}$&0.96\\ \hline\hline
  \end{tabular}
\caption{The comparison of the superconducting carriers density at $T=0$ with the density
of thermally activated electrons at $T=T_c$}
\label{nna}
\end{table}

\bigskip

From the data described above, we can obtain the condition of destruction of superconductivity, after heating  for  superconductors of type-I, as written in the equation:
\begin{equation}
{n_e}{({T_c})}\simeq 2{n_0}\label{nn}
\end{equation}

\section[The sound velocity]{The sound velocity of the zero-point oscillations condensate}
The wavelength of zero-point oscillations $\Lambda_0$  in this model is an analogue of the Pippard coherence length in the BCS.  As usually accepted \cite{Ketterson}, the coherence length $\xi = \frac{\hbar v_F} {4\Delta_0}$. The ratio of these lengths,  taking into account Eq.(\ref{TT0}), is simply the constant:
\begin{equation}
\frac{\Lambda_0}{\xi}\approx 8\pi^2\alpha^2\approx \cdot 10^{-3}.
\end{equation}

\vspace{0.5cm}

The attractive forces arising between the dipoles located at a distance $\frac{\Lambda_0}{2}$ from each other and vibrating in opposite phase, create pressure in the system:
\begin{equation}
P\simeq\frac{d\Delta_0}{dV}\simeq \frac{d_\Omega^2}{L_0^6}.
\end{equation}
In this regard, sound into this condensation should propagate with the velocity:
\begin{equation}
c_s\simeq\sqrt{ \frac{1}{2m_e}\frac{dP}{dn_0}}.
\end{equation}
After the appropriate substitutions, the speed of sound in the condensate can be expressed through the Fermi velocity of electron gas
\begin{equation}
c_s\simeq \sqrt{2\pi^2 \alpha^3}v_F\simeq 10^{-2} v_F.
\end{equation}

The condensate particles moving with velocity $c_S$ have the kinetic energy:
\begin{equation}
{2m_e c_s^2}\simeq\Delta_0.\label{cs}
\end{equation}
Therefore, by either heating the condensate  to the critical temperature when each of its volume obtains the energy $\mathcal{E}\approx n_0\Delta_0$, or initiating the current of its particles with a velocity exceeding $c_S$, can achieve the destruction of the condensate.
(Because the condensate of charged particles oscillations is considered, destroying its coherence can be also obtained at the application of a sufficiently strong magnetic field. See below.)

\section{The relationship $\Delta_0/kT_c$}\label{Delta-TcE}
From Eq.(\ref{nn}) and taking into account Eqs.(\ref{tcc}),(\ref{net})  and (\ref{n-0}), which were obtained for condensate,  we have:
\begin{equation}
\frac{\Delta_0}{kT_c}\simeq 1.86 .
\end{equation}

 This estimation of the relationship  $\Delta_0/kT_c$ obtained for condensate has a satisfactory agreement with the measured data \cite{Pool}, for type-I superconductors as listed in Table (\ref{DeTe}).\footnote{In the BCS-theory  $\frac{\Delta_0}{kT_c}\simeq1.76$.}

\bigskip

\begin{table}
\centering
\begin{tabular}{||c|c|c|c||}\hline\hline
&&&\\
  superconductor&$T_c$,K&$\Delta_0$,mev&$\frac{\Delta_0}{kT_c}$\\
  &&&\\\hline
  Cd &0.51&0.072&1.64\\
  Zn &0.85&0.13&1.77\\
  Ga &1.09&0.169&1.80\\
  Tl &2.39&0.369&1.79\\
  In &3.41&0.541&1.84\\
  Sn &3.72&0.593&1.85\\
  Hg &4.15&0.824&2.29\\
  Pb &7.19&1.38&2.22\\ \hline\hline
\end{tabular}
\caption{The value of ratio $\Delta_0/kT_c$ obtained experimentally for type-I superconductors}
\label{DeTe}
\end{table}

\chapter{Another superconductors}
\section{About type-II superconductors}{The estimation of properties of type-II superconductors}

In the case of type-II superconductors the situation is more complicated.

In this case, measurements show that these metals have an electronic specific heat that has an order of value greater than those calculated on the base of free electron gas model.

 The peculiarity of these metals is associated with the specific structure of their ions.
  They are transition metals with unfilled inner d-shell (see Table \ref{mII}).

It can be assumed that the increase in the electronic specific heat of these metals should be associated with a characteristic interaction of free electrons with the electrons of the unfilled d-shell.

\bigskip

\begin{table}
\centering
\begin{tabular}{||c|c||}\hline\hline
  superconductors &electron shells\\\hline
  $Ti$ &$3d^2 ~4s^2$\\
  $V$ &$3d^3  ~4s^2$\\
  $Zr$ &$4d^2~ 5s^2$\\
  $Nb$ &$4d^3 ~5s^2$\\
  $Mo$ &$4d^4~ 5s^2$\\
  $Tc$ &$4d^5 ~5s^2$\\
  $Ru$ &$4d^6~ 5s^2$\\
  $La$ &$5d^1 ~6s^2$\\
  $Hf$ &$5d^2 ~6s^2$\\
  $Ta$ &$5d^3 ~6s^2$\\
  $W$ &$5d^4~6s^2$\\
  $Re$ &$5d^5 ~6s^2$\\
  $Os$ &$5d^6 ~6s^2$\\
  $Ir$ &$5d^7~6s^2$\\\hline\hline
\end{tabular}
\caption{The external electron shells of elementary type-II superconductors}
\label{mII}
\end{table}

\bigskip

Since the heat capacity of the ionic lattice of metals is negligible at low temperatures,
only the electronic subsystem is thermally active .

At $T = 0$ the superconducting careers populates the energetic level $\mathcal{E}_F-\Delta_0$.
During the  destruction of superconductivity through heating, an each heated career increases its thermal vibration.
If the effective velocity  of vibration is $v_t$, its kinetic energy:
\begin{equation}
\mathcal{E}_k=\frac{mv_t^2}{2}\simeq \Delta_0
\end{equation}

Only a fraction of the heat energy transferred to the metal is consumed in order to increase the kinetic energy of the electron gas in the transition metals.

Another part of the energy will be spent on the magnetic interaction of a moving electron.

At contact with the d-shell electron, a freely moving electron  induces onto it the magnetic field of the order of value:
\begin{equation}
H\approx \frac{e}{r_c^2}\frac{v}{c}.
\end{equation}
The magnetic moment of d-electron is approximately equal to the Bohr magneton.
Therefore the energy of the magnetic interaction between a moving electron of conductivity and a d-electron is approximately equal to:
\begin{equation}
\mathcal{E}_\mu \approx \frac{e^2}{2r_c}\frac{v}{c}.
\end{equation}
This energy is not connected with the process of destruction of superconductivity.

Whereas, in metals with a filled d-shell (type-I superconductors), the whole heating energy  increases the kinetic energy of  the conductivity electrons and only a small part of the heating energy is spent on it in transition metals:
\begin{equation}
\frac{\mathcal{E}_k}{\mathcal{E}_\mu+\mathcal{E}_k}\simeq \frac{m v_t}{h}a_B.\label{kmuq}
\end{equation}
So approximately
\begin{equation}
\frac{\mathcal{E}_k}{\mathcal{E}_\mu+\mathcal{E}_k}\simeq \frac{a_B}{L_0}.\label{kmu}
\end{equation}

Therefore, whereas the dependence of the gap in type-I superconductors  from  the heat capacity is defined by Eq.(\ref{tcc}), it is necessary to  take into account the relation Eq.(\ref{kmu}) in type-II superconductors for the determination of this gap dependence.
As a result of this estimation, we can obtain:
\begin{equation}
\Delta_0\simeq \Theta \gamma^2\left(\frac{\mathcal{E}_k}{\mathcal{E}_\mu+\mathcal{E}_k}\right)\simeq \Theta \gamma^2\left(\frac{a_B}{L_0}\right)\frac{1}{2},\label{delta2}
\end{equation}
where $1/2$  is the fitting parameter.

 The comparison of the results of these calculations with the measurement data (Fig.(\ref{tc2g})) shows that for the majority of type-II superconductors the estimation Eq.(\ref{delta2}) can be considered quite satisfactory.\footnote{The lowest critical temperature was measured for Mg. It is approximately equal to 1mK.  Mg-atoms in the metallic state are given two electrons into  the electron gas of conductivity. It is confirmed by the fact that the pairing of these electrons, which manifests itself in the measured value of the flux quantum \cite{Sharv}, is observed above $T_c$.
 It would seem that in view of this metallic Mg-ion must have electron shell like the Ne-atom.
 Therefore it is logical to expect that the critical temperature of Mg can be calculated by the formula for I-type  superconductors. But actually in order to get the value of $T_c\approx 1mK$,  the critical temperature of Mg should be calculated by the formula (\ref {delta2}), which is applicable to the description of metals with an unfilled inner shell. This suggests that the ionic core of magnesium metal apparently is not as simple as the completely filled Ne-shell.}

\begin{figure}
\includegraphics[scale=.5]{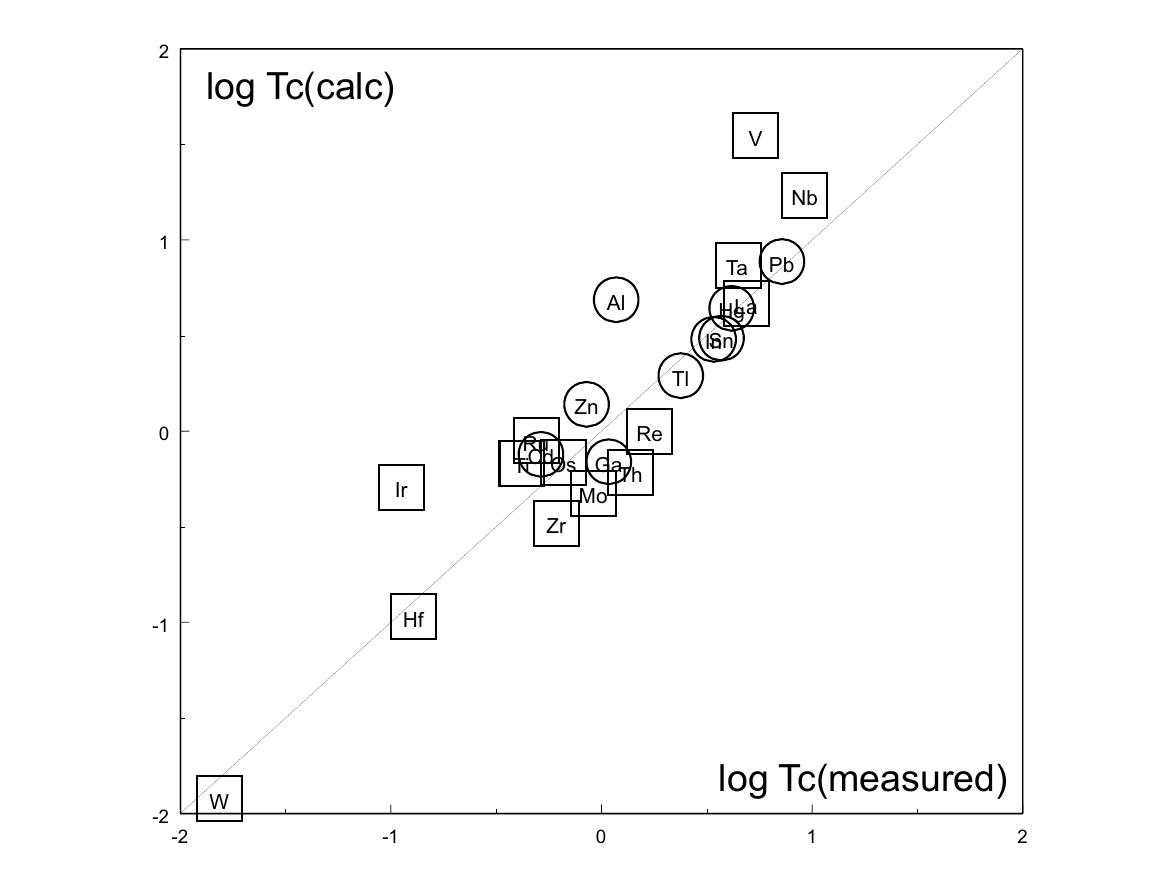}
\caption{The comparison of the calculated values of critical temperatures of superconductors with measurement data.
Circles  relate to type-I superconductors, squares show type-II superconductors.
On the abscissa, the measured values of critical temperatures are plotted, on ordinate, the calculated estimations are plotted. The calculations of critical temperatures for type-I superconductors were made by using Eq.(\ref{tcc}) and
the estimations for type-II superconductors was obtained by using Eq.(\ref{delta2}).}\label{tc2g}
\end{figure}


\section{Alloys and high-temperature superconductors}
In order to understand the mechanism of high temperature superconductivity, it is important to establish whether the high-$T_c$ ceramics are the I or II-type superconductors, or whether they are a special class of superconductors.

 In order to determine this, we need to look at the above established dependence of critical parameters from the electronic specific heat and also consider that the specific heat of superconductors I and II-types are differing considerably.

There are some difficulties by determining the answer this way: as we  do not precisely know the density of the electron gas in high-temperature superconductors.
However, the densities of atoms in metals do not differ too much and we can use Eq.(\ref{tcc}) for  the solution of the problem  of the I- and II-types superconductors  distinguishing.

If parameters of type-I superconductors are inserted into this equation, we obtain quite a satisfactory estimation of the critical temperature (as was done above, see Fig.\ref{tc2g}). For the type-II superconductors` values, this  assessment gives an overestimated value due to the fact that type-II superconductors' specific heat has additional term associated with the magnetization of d-electrons.

This analysis therefore, illustrates a possibility where we can divide all superconductors into two groups, as is evident from the Fig.(\ref{gamma2}).

\begin{figure}
\hspace{1.5cm}
\includegraphics[scale=.4]{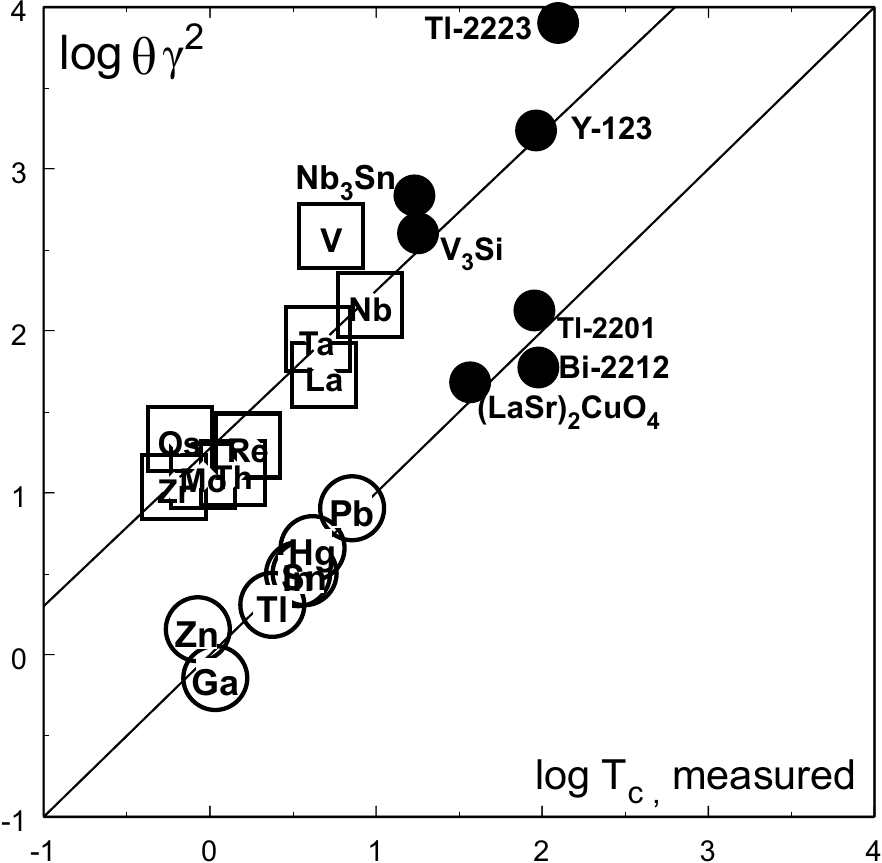}
\caption {The comparison of the calculated parameter $\Theta\gamma^2$ with the measurement of the critical temperatures of elementary superconductors and some superconducting compounds.}\label{gamma2}
\end{figure}

It is generally assumed that we  consider alloys $Nb_3Sn$ and $V_3Si$ as the type-II superconductors. This assumption seems quite normal because they are placed in close surroundings of Nb.
Some excess of the calculated critical temperature over the experimentally measured value for ceramics $Ta_2Ba_2Ca_2Cu_3O_{10}$ can be attributed to  the measured heat capacity that may have been created by not only conductive electrons, but also non-superconducting elements (layers) of ceramics. It is already known that it, as well as ceramics $YBa_2Cu_3O_7$, belongs to the type-II superconductors. However, ceramics (LaSr)$_2$Cu$_4$, Bi-2212 and Tl-2201, according to this figure should be regarded as type-I superconductors, which is unusual.


\chapter{About the London penetration depth}
\label{LondonE}

\section{The magnetic energy of a moving electron}
To avoid these incorrect results, let us consider a balance of magnetic energy in a superconductor within magnetic field.
This magnetic energy is composed of energy from a penetrating external magnetic field and magnetic energy of  moving electrons.


By using formulas \cite{Bek}, let us estimate the ratio of the magnetic and kinetic energy of an electron (the charge of $e$ and the mass $m_e$) when it moves rectilinearly with a velocity $v\ll c$.

The density of the electromagnetic field momentum  is expressed by the equation:
\begin{equation}
\mathbf{g}=\frac{1}{4\pi c}[\mathbf{E}\mathbf{H}]
\end{equation}

While moving with a velocity $\mathbf{v}$, the electric charge carrying the electric field with intensity $E$ creates a magnetic field
\begin{equation}
\mathbf{H}=\frac{1}{c}[\mathbf{E}\mathbf{v}]\label{Hvc}
\end{equation}
with the density of the electromagnetic field momentum (at $v\ll c$)
\begin{equation}
\mathbf{g}=\frac{1}{4\pi c^2}[\mathbf{E}[\mathbf{v}\mathbf{E}]]=\frac{1}{4\pi c^2}\left(\mathbf{v}E^2-\mathbf{E}(\mathbf{v}\cdot \mathbf{E})\right)
\end{equation}
As a result, the momentum of the electromagnetic field of a moving electron
\begin{equation}
\mathbf{G}=\int_V \mathbf{g}dV=
\frac{1}{4\pi c^2}\left(\mathbf{v}\int_V E^2 ~dV - \int_V \mathbf{E}~E~v~ cos\vartheta ~dV \right)\label{cos}
\end{equation}
The integrals are taken over the entire space, which is occupied by particle fields, and $\vartheta$ is the angle between the particle velocity and the radius vector of the observation point. By calculating the last integral in the condition of the axial symmetry with respect to $\mathbf{v}$, the contributions from the components of the vector $\mathbf{E}$, which is perpendicular to the velocity, cancel each other for all pairs of elements of the space (if they located diametrically opposite on the magnetic force line). Therefore, according to  Eq.(\ref{cos}), the component of the field which is collinear to $\mathbf{v}$
\begin{equation}
\frac{E~cos\vartheta \cdot \mathbf{v}}{v}
\end{equation}
can be taken instead of the vector $\mathbf{E}$.
By taking this information into account, going over to the spherical coordinates and integrating over angles, we can obtain
\begin{equation}
\mathbf{G}=
\frac{\mathbf{v}}{4\pi c^2}\int_{r}^\infty E^2\cdot 4\pi r^2~dr
\end{equation}
If we limit the integration of the field by the Compton electron radius $r_C=\frac{\hbar}{m_e c}$, ~ \footnote{Such effects as the pair generation force us to consider the radius of the "quantum electron" \ as approximately equal to Compton radius \cite{Messia}.} then $v\ll c$, and we obtain:
\begin{equation}
\mathbf{G}=\frac{\mathbf{v}}{4\pi c^2}\int_{r_C}^\infty E^2\cdot 4\pi r^2~dr=
\frac{\mathbf{v}}{c^2} \frac{e^2}{r_C}.
\end{equation}
In this case by taking into account Eq.(\ref{Hvc}), the magnetic energy of a slowly moving electron pair is equal to:
\begin{equation}
\mathcal{E}=\frac{{v}{G}}{2}=
\frac{{v^2}}{c^2} \frac{e^2}{2r_C}=\alpha\frac{m_e v^2}{2}.\label{EEalfa}
\end{equation}

\section{The magnetic energy and the London penetration depth}
\label{London-new2E}
The energy of external magnetic field into volume $dv$:
\begin{equation}
\mathcal{E}=\frac{H^2}{8\pi}dv.
\end{equation}
At a density of superconducting carriers $n_s$, their magnetic energy per unit volume in accordance with (\ref {EEalfa}):
\begin{equation}
\mathcal{E}_H\simeq\alpha n_s\frac{m_2 v^2}{2}=\alpha\frac{m_e j_s^2}{2n_s e},
\end{equation}
where $j_s=2e n_s v_s$ is the density  of a current of superconducting carriers.

Taking into account the Maxwell equation
\begin{equation}
\mathbf{rot H}=\frac{4\pi}{c}\mathbf{j}_s,
\end{equation}
the magnetic energy of moving carriers can be written as
\begin{equation}
\mathcal{E}_H\simeq \frac{\widetilde{\Lambda}^2}{8\pi}(rot H)^2,
\end{equation}
where we introduce the notation
\begin{equation}
\widetilde{\Lambda}=\sqrt{\alpha \frac{m_e c^2}{4\pi n_s e^2}}=\sqrt{\alpha}\Lambda_L.\label{lam}
\end{equation}
In this case, part of the free energy of the superconductor connected with the application of a magnetic field is equal to:
\begin{equation}
\mathcal{F}_H=\frac{1}{8\pi}\int_V\left(H^2+\widetilde{\Lambda}^2(rot H)^2\right)dv.
\end{equation}
At the minimization of the free energy, after some simple transformations we obtain
\begin{equation}
\mathbf{H}+\widetilde{\Lambda}^2 \mathbf{rot rot H}=0,
\end{equation}
thus  $\widetilde{\Lambda}$ is the depth of magnetic field penetration into the superconductor.

In view of Eq.(\ref{n-0}) from Eq.(\ref{lam}) we can estimate the values of London penetration depth (see table (\ref{London2})).
The consent of the obtained values with the measurement data can be considered quite satisfactory.

\vspace{0.5cm}

\begin{table}
\centering
\begin{tabular}{||c|c|c|c||} \hline\hline
&&&\\
super-&$\lambda_L$,$10^{-6}$cm &$\widetilde{\Lambda}$,$10^{-6}$cm&\\
conductors     & measured \cite{Linton} & calculated&$\widetilde{\Lambda}/\lambda_L$\\
 &                         &Eq.(\ref{lam})&\\\hline
  Tl  &    9.2   &     11.0&1.2 \\
  In  &    6.4   &     8.4&1.3 \\
  Sn  &    5.1   &     7.9&1.5 \\
  Hg  &    4.2   &     7.2&1.7 \\
  Pb  &    3.9   &     4.8&1.2 \\\hline\hline
\end{tabular}
\caption{Corrected values of London penetration depth}
\label{London2}
\end{table}

\vspace{0.5cm}

The resulting refinement may be important for estimates within the frame of Ginzburg-Landau theory, where the London
 penetration depth is used as a comparison of calculations and specific parameters of superconductors.


\vspace{0.5cm}
\newpage
\chapter{Three words to experimenters}

{\it{

 The history of the Medes is obscure and incomprehensible.

 Scientists divide it, however, into three periods:

 The first is the period,  which is absolutely unknown.

 The second is one which is followed after the first.

 And finally, the third period is a period which is known

 to the same degree as two firsts.}}\\

\hspace{2cm}A. Averchenko $\ll$The  World History$\gg$

 \hspace{2cm} in Ja. Zeldovich presentation.

\section[Is the room-temperature superconductivity possible?]{Why creation of room-temperature superconductors are hardly probably?}
The understanding of the mechanism of the superconducting state should open a way towards finding a solution to the technological problem. This problem was just a dream in the last century:
 the dream to create a superconductor that would be easily  produced (in the sense of ductility) and had  high critical temperature.

In order to move towards achieving this goal,
it is important firstly to understand the mechanism that limits the  critical properties of superconductors.

Let us consider a superconductor with a large limiting current.
The length of their localisation determines the limiting  momentum of  superconducting carriers:
\begin{equation}
p_c\simeq\frac{2\pi \hbar}{L_0}.
\end{equation}
Therefore, by using  Eq.(\ref{cs}), we can compare the critical velocity of superconducting carriers with the sound velocity:
\begin{equation}
v_c=\frac{p_c}{2m_e}\simeq c_s
\end{equation}
 and both these velocities are about a hundred times smaller than the Fermi velocity.

The sound velocity in the crystal lattice of  metal $v_{s}$, in accordance with the Bohm-Staver relation \cite{Ashkr}, has  approximately the same value:
\begin{equation}
v_s\simeq \frac{k T_D}{E_F}v_F\simeq 10^{-2}{v_F}.
\end{equation}
This therefore, makes it possible to consider superconductivity being destroyed as a superconducting carrier overcomes
the sound barrier.
That is, if they moved without friction at a speed that was less than that of sound, after it gained speed and the speed of sound was surpassed,
it then acquire a mechanism of friction.

 Therefore, it is conceivable that if the speed of sound in the metal lattice $v_s<c_s$, then it would create a restriction on the limiting current in superconductor.

 If this is correct, then superconductors with high critical parameters should have not only a high Fermi energy of their electron gas, but also a high speed of sound in their lattice.

It is in agreement with the fact that  ceramics have higher elastic moduli compared to metals and alloys, and also posses  much higher critical temperatures (Fig.{\ref{vsg}}).

\begin{figure}
\hspace{-0.5cm}
\includegraphics[scale=0.4]{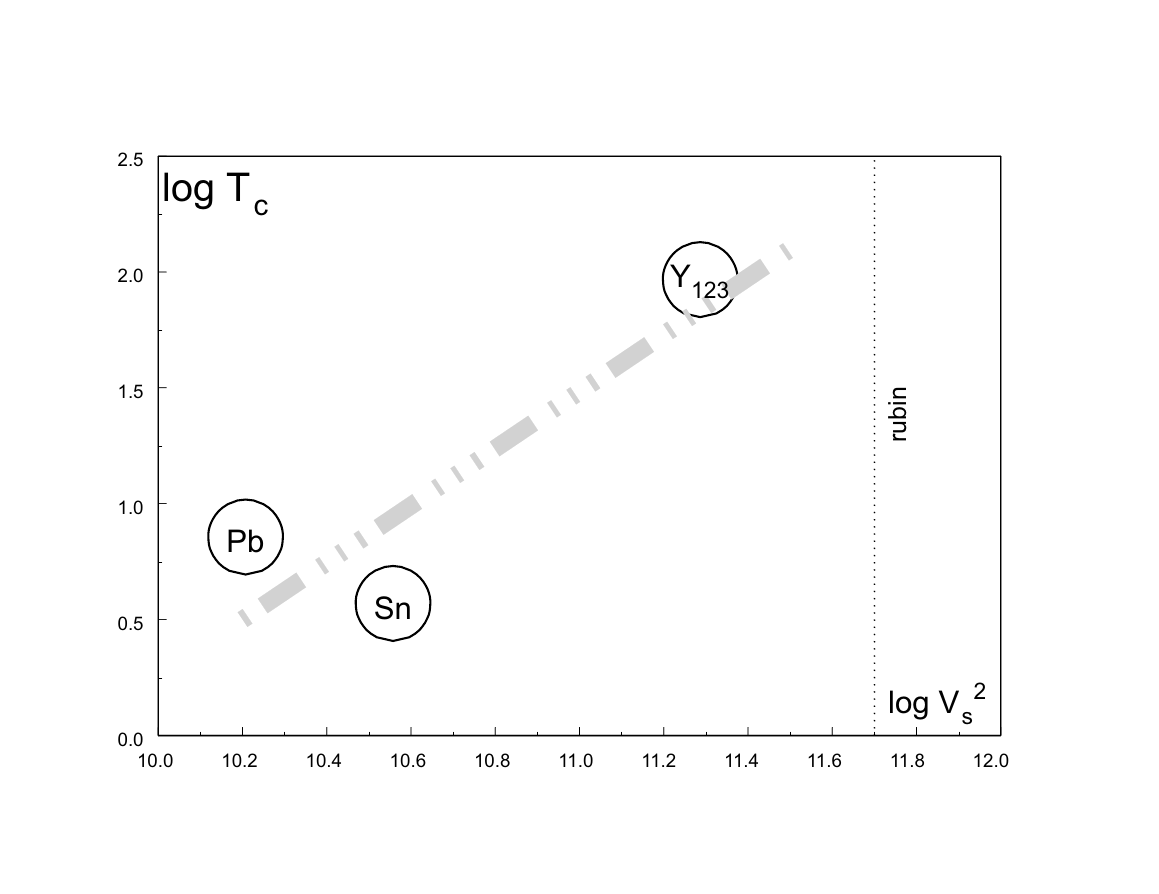}
\caption{The schematic representation of the dependence of critical temperature on the speed of sound in superconductors. On the ordinate, the logarithm of the superconductor's critical temperature is shown. On the abscissa, the logarithm of the square of the speed of sound  is shown (for Sn and Pb - the transverse velocity of sound is shown, because it is smaller).  The speed of sound in a film   was used for yttrium-123 ceramics. The dashed line shows the value of the transverse velocity of sound in sapphire, as some estimation of the limit of its value.
It  can be seen that this estimation leads to the restriction on the critical temperature in the range about $0^o$C - the dot-dashed line.}\label{vsg}
\end{figure}

\vspace{1cm}

The dependence of the critical temperature on the square of the speed of sound \cite{Gol} is illustrated in Fig.(\ref{vsg}).

  This figure, which can be viewed only as a rough estimation due to the lack of necessary experimental data, shows that the elastic modulus of ceramics with a critical temperature close to the room temperature, should be close to the elastic modulus of sapphire, which is very difficult to achieve.

  In addition, such ceramics would be deprived of yet an other important quality - their adaptability.
Indeed, in order to obtain a thin wire, we  require a plastic superconductor.

A solution of this problem would be to find  a material that possesses an acceptably high critical temperature (above 80K) and also  experiences a phase transition at an even higher temperature of heat treatment.
It would be possible to make a thin wire from a superconductor near the point of phase transition, as the elastic modules  are typically  not usually very strong at this stage.

\section{Magnetic electron pairing}

This considered formation of mechanism for the superconducting state provides a possibility of obtaining the estimations  of the critical parameters of superconductors, which in most cases is in satisfactory agreement with measured data. For some superconductors, this agreement is stronger, and for other, such as Ir, Al, V (see Fig.(\ref{tc2g})), it is expedient to carry out further theoretical and experimental studies due to causes of deviations.

The mechanism of magnetic electron pairing is also of fundamental interest in order to further clarify this.

As was found earlier, in the cylinders made from  certain superconducting metals (Al\cite{Shab} and Mg\cite {Sharv}), the observed magnetic flux quantization has exactly the same period above $T_c$ and that below $T_c$. The authors of these studies attributed this to the influence of a special effect. It seems more natural to think that the stability of the  period is a result of the pairing of electrons due to magnetic dipole-dipole interaction continuing to exist at temperatures above $T_c$, despite the disappearance of the material's superconducting properties. At this temperature  the coherence of the zero-point fluctuations is destroyed, and with it so is the superconductivity.

The pairing of electrons due to dipole-dipole interaction should  be absent in the monovalent metals. In these metals, the   conduction electrons are localized in the lattice at very large distances from each other.

It is therefore interesting to compare the  period of quantization in the two cases.
In a thin cylinder made of a superconductor, such as Mg, above $T_c$ the quantization period is equal to $\frac{2\pi\hbar c}{2e}$. In the same cylinder of a noble metal (such as gold), the sampling period should be twice as large.

\section[The effect of isotopic substitution]{The effect of isotopic substitution on the condensation of zero-point oscillations}
The attention of experimentalists could be attracted
 to the isotope effect in superconductors, which served as a starting point of the B-BCS theory.
 In the '50s, it had been experimentally established that there is a dependence of the critical temperature of superconductors due to the mass of the isotope. As the effect depends on the ionic mass, this is considered to be due to  the fact that it is based on the vibrational (phonon) process.

The isotope effect for a number of I-type superconductors - $Zn, Sn, In, Hg, Pb $ - can be described by the relationship:
\begin{equation}
\sqrt{M_i}T_c=const,\label{ise}
\end{equation}
where $M_i$ is the mass of the isotope, $T_c$ is the critical temperature.
The isotope effect in other superconductors can either be described by other dependencies,
 or is absent altogether.

In recent decades, however, the effects associated with the replacement of isotopes in the metal lattice have been studied in detail.
It was shown that the zero-point oscillations of  ions in the lattice of many metals are non-harmonical.
Therefore, the isotopic substitution can directly affect the lattice parameters, the density of the lattice and the density of the electron gas in the metal, on its Fermi energy  and on other  properties of the electronic subsystem.

The direct study of the effect of isotopic substitution on the lattice parameters of superconducting metals has not been carried out.

The results of measurements made on $Ge$, $Si$, diamond and light metals, such as $Li$ \cite{Inyu}, \cite{Kogan} (researchers prefer to study crystals, where the isotope effects are large, and it is easier to carry out appropriate measurements), show that there is square-root dependence of the force constants on the isotope mass, which was required by Eq.(\ref{ise}). The same dependence of the force constants on the mass of the isotope has been found in tin \cite{Wang}.

Unfortunately, no direct experiments of the effect of isotopic substitution on the electronic properties (such as the electronic specific heat and the Fermi energy), exist for metals substantial for our consideration.

Let us consider what should be expected in such measurements. A convenient choice for the superconductor is mercury, as it has many isotopes and their isotope effect has been carefully measured back in the 50s of the last century as aforementioned.

The linear dependence of the critical temperature of a superconductor on its Fermi energy (Eq.(\ref{TT0})) and also the existence  of the isotope effect suggests the dependence of the ion density in the  crystal lattice from the mass of the isotope. Let us consider what should be expected in such measurements.

Even then, it was found that the isotope effect is described by Eq.(\ref{ise}) in only a few superconductors. In others, it displays different values, and therefore in a general case it can be described by introducing of the parameter
$\mathfrak{a}$:
\begin{equation}
M_i^\mathfrak{a} T_c=Const.
\end{equation}
At taking into account Eq.(\ref{TT0}), we can write
\begin{equation}
T_c\sim \mathcal{E}_F\sim n_e^{2/3}
\end{equation}

The parameter $l$ which characterizes the ion lattice obtains an increment
$\Delta l$ with an isotope substitution:
\begin{equation}
\frac{\Delta l}{l} = -\frac{\mathfrak{a}}{2}\cdot \frac{\Delta M_i}{M_i},
\end{equation}
where ${M_i}$ and $\Delta M_i$ are the mass of isotope and its increment.

It is generally accepted that in an accordance with the terms of the phonon mechanism, the parameter $\mathfrak{a}\approx \frac{1}{2}$ for mercury.
 However, the analysis of experimental data \cite{Maxwell}-\cite{Serin} (see Fig.(\ref{Hgg})) shows that this parameter is actually closer to $1/3$. Accordingly, one can expect that the  ratio of the mercury parameters is close to:
\begin{equation}
\frac{\left(\frac{\Delta l}{l}\right)}{\left(\frac{\Delta M_i}{ M_i}\right)}\approx -\frac{1}{6} .
\end{equation}

\newpage

\chapter[Superfluidity]{Superfluidity as a sequence of  ordering of zero-point oscillations}
\label{HeE}
\section{Zero-point oscillations of He atoms and superfluidity}

The main features of superfluidity of liquid helium became clear few decades ago
\cite{Landau}, \cite{Halat}. L.D.Landau explains this phenomenon as the manifestation of a quantum behavior of the macroscopic object.

However, the causes and mechanism of the formation of superfluidity  are not clear till our days.
There is no explanation why the $\lambda$-transition in helium-4 occurs at about 2 K, that is  about twice less than   its boiling point:
\begin{equation}
\frac{T_{boiling}}{T_\lambda}\approx 1.94,\label{f022}
\end{equation}
while for helium-3, this transition is observed only at temperatures  about a thousand times smaller.

 The related phenomenon, superconductivity, can be regarded as superfluidity of a charged liquid. It can be quantitatively described considering it as the consequence of ordering of zero-point oscillations of electron gas. Therefore it seems appropriate to consider superfluidity from the same point of view.

Atoms in liquid helium-4 are electrically neutral, as they have no dipole moments and do not form molecules.
  Yet some electromagnetic mechanism should be responsible for phase transformations of liquid helium (as well as in other condensed substance where phase transformations are related to the changes of energy of the same scale).

 F. London has demonstrated already  in the 1930's \cite{FLondon}, that there is an interaction between  atoms in the ground state, and this interaction is of a quantum nature. It can be considered as a kind of  the Van-der-Waals interaction.
 Atoms in their ground state (T = 0) perform zero-point oscillations. F.London was considering  vibrating atoms as three-dimensional oscillating dipoles which  are connected to each other by the electromagnetic interaction. He proposed the name the dispersion interaction for this interaction of atoms in the ground state.

\section{The dispersion effect in interaction of atoms in the ground state}
Following F.London \cite{FLondon}, let us consider two spherically symmetric atoms without non-zero average dipole moments.  Let us suppose that at some time the charges of these atoms are  fluctuationally displaced from the equilibrium states:
\begin{equation}
r_1=(x_1,y_1,z_1)
\end{equation}
\begin{equation}
r_{2}=(x_{2},y_{2},z_{2})\nonumber
\end{equation}
If atoms are located along the Z-axis at the distance $L$ of each other, their potential energy can be written as:
\begin{equation}
\begin{array}{l}
\mathcal{H}=\underbrace{\frac{e^2 r_1^2}{2a}+\frac{e^2 r_{2}^2}{2a}}_{elastic~dipoles~energy}+\\
\qquad \qquad \qquad +\underbrace{\frac{e^2}{L^3}(x_1 x_{2}+y_1y_{2}-2z_1z_{2})}_{elastic~dipoles~interaction}.\label{q0}
\end{array}
\end{equation}
where $a$ is the atom polarizability.

The Hamiltonian can be diagonalized by using the normal coordinates of symmetric and antisymmetric displacements:
\begin{displaymath}
r_s\equiv
\left\{
\begin{array}{ll}
x_s=\frac{1}{\sqrt{2}}(x_1+x_2) \\
y_s=\frac{1}{\sqrt{2}}(y_1+y_2) \\
z_s=\frac{1}{\sqrt{2}}(z_1+z_2) \\
\end{array}
\right.
\end{displaymath}
and
\begin{displaymath}
r_a\equiv
\left\{
\begin{array}{ll}
x_a=\frac{1}{\sqrt{2}}(x_1-x_2) \\
y_a=\frac{1}{\sqrt{2}}(y_1-y_2) \\
z_a=\frac{1}{\sqrt{2}}(z_1-z_2) \\
\end{array}
\right.
\end{displaymath}
This yields
\begin{displaymath}
\begin{array}{ll}
x_1=\frac{1}{\sqrt{2}}(x_s+x_a) \\
y_1=\frac{1}{\sqrt{2}}(y_s+y_a) \\
z_1=\frac{1}{\sqrt{2}}(z_s+z_a) \\
\end{array}
\end{displaymath}
and
\begin{displaymath}
\begin{array}{ll}
x_2=\frac{1}{\sqrt{2}}(x_s-x_a) \\
y_2=\frac{1}{\sqrt{2}}(y_s-y_a) \\
z_2=\frac{1}{\sqrt{2}}(z_s-z_a) \\
\end{array}
\end{displaymath}
As the result of this change of variables we obtain:
{{
\begin{equation}
\begin{array}{l}
\mathcal{H}=\frac{e^2}{2a}(r_s^2+r_a^2)+\\
+\frac{e^2}{2L^3}(x_s^2+y_s^3-2z_s^2-x_a^2-y_a^2+2z_a^2)= \\
=\frac{e^2}{2a}\bigg[\left(1+\frac{a}{L^3}\right)(x_s^2+y_s^2) + \\
\qquad\qquad+\left(1-\frac{a}{L^3}\right)(x_a^2+y_a^2)+ \\
\qquad\qquad+\left(1-2\frac{a}{L^3}\right)z_s^2 +
\left(1+2\frac{a}{L^3}\right)z_a^2\bigg].\label{q01}
\end{array}
\end{equation}
}}
Consequently, frequencies of oscillators depend on their  orientation and they are determined by the equations:
{
\begin{equation}
\begin{array}{l}
\Omega_{0x}^{s\atop a}=\Omega_{0y}^{s\atop a}=\Omega_0\sqrt{1\pm\frac{a}{L^3}}\approx\\
\approx\Omega_0\left({1\pm\frac{a}{L^3}}-\frac{a^2}{8L^6}\pm...\right),  \\
\\
\Omega_{0z}^{s\atop a}=\Omega_0\sqrt{1\mp\frac{2a}{L^3}}\approx\\
\approx\Omega_0\left({1\mp\frac{a}{L^3}}-\frac{a^2}{2L^6}\mp...\right),
\end{array}
\end{equation}
}
where
\begin{equation}
\Omega_0=\frac{2\pi e}{\sqrt{ma}}
\end{equation}
is natural frequency of the electronic shell of the atom (at $L\rightarrow\infty$).
The energy of zero-point oscillations is
\begin{equation}
\mathcal{E}=\frac{1}{2}\hbar(\Omega_0^s+\Omega_0^a)
\end{equation}
It is easy to see that the description of interactions between neutral atoms do not contain terms $\frac{1}{L^3}$, which are characteristics for the interaction of zero-point oscillations in the electron gas (Eq.(\ref{Lz3})) and
which are responsible for the occurrence of superconductivity. \\
The terms that are proportional to $\frac{1}{L^6}$ manifest themselves in interactions of neutral atoms.\\

It is important to emphasize that the energies of interaction are different for  different orientations of zero-point oscillations. So the interaction of zero-point oscillations oriented along the direction connecting the atoms leads to their attraction with energy:
\begin{equation}
\mathcal{E}_z = - \frac{1}{2}\hbar\Omega_0\frac{A^2}{L^6},\label{f1}
\end{equation}
while the sum energy of the attraction of the oscillators of the perpendicular directions (x and y)  is equal to one half of it:
\begin{equation}
\mathcal{E}_{x+y} = - \frac{1}{4}\hbar\Omega_0\frac{A^2}{L^6}\label{f2}
\end{equation}
(the minus sign is taken here because  for this case   the opposite direction of dipoles  is energetically favorable).

\vspace{1cm}

\section[The main characteristics of  superfluid helium]{The estimation of main characteristic parameters of  superfluid helium}
\subsection{The main characteristic parameters of the zero-point oscillations of atoms in superfluid helium-4}
$\\$
There is no repulsion in a gas of neutral bosons.
Therefore, due to attraction between the atoms at temperatures below
\begin{equation}
T_{boil}=\frac{2}{3k}\mathcal{E}_z
\end{equation}
this gas collapses and a liquid  forms.

At twice lower temperature
\begin{equation}
T_\lambda=\frac{2}{3k}\mathcal{E}_{x+y}
\end{equation}
all zero-point oscillations become ordered. It creates an additional attraction and forms a single quantum ensemble.

A density of the boson condensate  is limited by  zero-point oscillations  of its atoms.
At condensation the distances between the atoms become approximately equal  to  amplitudes of zero-point oscillations.\\

Coming from it, we can calculate the basic properties of an ensemble of atoms with ordered zero-point oscillations, and compare them with measurement properties of superfluid helium.

We can assume that the radius of a helium atom is equal to the Bohr radius $a_B$, as it follows from quantum-mechanical calculations.
Therefore, the energy of electrons on the s-shell of this atom can be considered to be equal:
\begin{equation}
\hbar\Omega_0 = \frac{4e^2}{a_B}
\end{equation}
As the polarizability of atom is approximately equal to its volume \cite{Fr}
\begin{equation}
A\simeq a_B^3,\label{vA}
\end{equation}
the potential energy of dispersive interaction (\ref{f2}), which causes the ordering zero-point oscillations in the ensemble of atoms, we can represent by the equation:
\begin{equation}
\mathcal{E}_{x+y} = - \frac{e^2}{a_B}a_B^6 n^2,\label{W}
\end{equation}
where the density of helium atoms
\begin{equation}
n=\frac{1}{L^3}
\end{equation}

\subsubsection{The velocity of zero-point oscillations of helium atom}
It is naturally  to suppose that zero-point oscillations of atoms are  harmonic  and the equality of kinetic and potential energies are characteristic for them:
\begin{equation}
\frac{M_4 \widehat{v_0}^2}{2} - \frac{e^2}{a_B}a_B^6 n^2 = 0,
\end{equation}
where $M_4$ is mass of helium atom, $\widehat{v_0}$ is their averaged velocity of harmonic zero-point oscillations.\\

Hence, after simple transformations we obtain:
\begin{equation}
\widehat{v_0}=c\alpha^3\left\{\frac{n}{n_0}\right\},\label{Ea31}
\end{equation}
where the notation is introduced:
\begin{equation}
n_0=\frac{\alpha^2}{a_B^3}\sqrt{\frac{M_4}{2m_e}}.\label{}
\end{equation}
If the expression in the curly brackets
\begin{equation}
\frac{n}{n_0}=1,\label{edin}
\end{equation}
we obtain
\begin{equation}
\widehat{v_0}=c\alpha^3\cong 116.5~
m/s.\label{alpha3}
\end{equation}

\subsubsection{The density of liquid helium}
The condition (\ref{edin}) can be considered as the definition of the density of helium atoms in the superfluid state:
\begin{equation}
n=n_0=\frac{\alpha^2}{a_B^3}\sqrt{\frac{M_4}{2m_e}}\cong 2.172\cdot 10^{22}~ atom/cm^3.\label{nnn}
\end{equation}
According to this definition, the density of liquid helium-4
\begin{equation}
\gamma_4 = n M_4\cong 0.1443~ g/cm^3
\end{equation}
that is in good agreement with the measured density of the liquid helium $0.145~g/cm^3$ for $T\simeq T_\lambda$.

Similar calculations for liquid helium-3 gives the density  $0.094~g/cm^3$, which can be regarded as consistent with its density $0.082 ~g/cm^3$ experimentally measured near the boiling point.

\subsubsection{The dielectric constant of liquid helium}
To estimate the dielectric constant of helium we can use
the Clausius-Mossotti equation \cite{Fr}:
\begin{equation}
\frac{\varepsilon-1}{\varepsilon+2}=\frac{4\pi}{3}{n}{A}.\label{KM}
\end{equation}
At taking into account Eq.(\ref{vA}), we obtain
\begin{equation}
\varepsilon\approx 1.040,
\end{equation}
that differs slightly from the dielectric constant of the liquid helium,
measured near the $\lambda$-point \cite{Russ}:
\begin{equation}
\varepsilon\approx 1.057
\end{equation}

\subsubsection{The temperature of $\lambda$-point}
The superfluidity is destroyed at the temperature $T_\lambda$, at which the energy of thermal motion is compared with the energy of the Van-der-Waals bond
in superfluid condensate
\begin{equation}
\frac{3}{2}kT_\lambda - \frac{e^2}{a_B}a_B^6 n^2 = 0.
\end{equation}
With taking into account Eq.(\ref{nnn})
\begin{equation}
T_{\lambda}=\frac{1}{3k}\frac{M_4}{m_e}\frac{\alpha^4 e^2}{a_B}\label{f021}
\end{equation}
or after appropriate substitutions
\begin{equation}
T_{\lambda}=\frac{1}{3}\frac{M_4 c^2 \alpha^6}{k} =2.177 K,\label{f022}
\end{equation}
that is in very good agreement with the measured value $T_\lambda = 2.172K$.\footnote{There is a unexpected fact. The expression (\ref{f022}) for the temperature of $\lambda$-transition is given without any explanations in some articles of Internet at citing of patents \cite{Il}.  These articles and patents say nothing at all about zero-point oscillations, and  don't give generally any explanations of the reasons that allowed to write this expression.}

\subsubsection{The boiling temperature of liquid helium}
After comparison of Eq.(\ref{f1}) - Eq.(\ref{f2}), we have
\begin{equation}
T_{boil}=2T_\lambda=4.35 K\label{Tboil}
\end{equation}
This is the basis for the assumption that the liquefaction of helium is due to the attractive forces between the atoms with ordered
lengthwise  components of their oscillations.

\subsubsection{The velocity of the first sound in liquid helium}
It  is known from the theory of the harmonic oscillator that the maximum value of its velocity  is twice bigger than its average velocity.
In this connection, at assumption that the first sound speed  $c_{s1}$ is limited by this maximum speed oscillator, we obtain
\begin{equation}
c_{s1}=2\widehat{v_0}\simeq 233~m/s.\label{v01}
\end{equation}
It is in consistent with the measured value of the velocity of the first sound in helium, which has the maximum value of $238.3~m/s$ at $T\rightarrow 0$  and decreases with increasing temperature up to about $220~m/s$ at $T=T_\lambda$.\\

The results obtained in this section  are summarized  for clarity in the Table.(\ref{DT}).\\

The measurement data in this table are mainly quoted by \cite{Kik} and \cite{Russ}.\\

{
{
\begin{table}
\centering
\begin{tabular}{||c|c|c||c||}\hline\hline
&&&\\
  &defining &calculated&measured\\
  parameter&&&\\
  &formula&value&value\\
  &&&\\\hline
  the velocity of zero-point&&&\\
  oscillations of&$\widehat{v_0}=c\alpha^3$&$116.5$~&\\
  helium atom &&m/s&\\\hline\hline
  The density of atoms&&&\\
  in liquid &$n=\sqrt{\frac{M_4}{2m_e}}\frac{\alpha^2}{a_B^3}$&$2.172\cdot 10^{22}$&\\
  helium &&$atom/cm^3$&\\\hline\hline
  The density&&&\\
  of liquid helium-4&$\gamma=M_4 n$&$144.3$&$145_{T\simeq T_{\lambda}}$\\
   $ g/l$&&&\\\hline\hline
  The dielectric&&&$1.048_{T\simeq 4.2}$\\
  constant&$\frac{\varepsilon-1}{\varepsilon+2}=\frac{4\pi}{3}{\alpha^2}{\sqrt{\frac{M_4}{2m_e}}}$&1.040&\\
  of liquid helium-4&&&$1.057_{T\simeq T_{\lambda}}$\\\hline\hline
  The temperature &&&\\
  &$T_\lambda\simeq\frac{M_4 c^2 \alpha^6}{3}$&$2.177$&$2.172$\\
  $\lambda$-point,K&&&\\\hline\hline
  The boiling &&&\\
  temperature&$T_{boil}\simeq 2T_\lambda$&$4.35$&$4.21$\\
  of helium-4,K&&&\\\hline\hline
  The first sound &&&\\
  velocity,&$c_{s1}=2\widehat{v_0}$&$233$&$238.3_{T\rightarrow 0}$\\
  $m/s$&&&\\\hline\hline
\end{tabular}
\caption{Comparison of the calculated values of liquid helium-4 with the measurement data}
\label{DT}
\end{table}
}}
\newpage
\subsection{The estimation of characteristic properties of He-3}
The results of similar calculations for the helium-3 properties are summarized in the Tab.(\ref{DT3}).
\bigskip

{
{\begin{table}
\centering
\hspace{-1cm}\begin{tabular}{||c|c|c||c||}\hline\hline
&&&\\
 &defining &calculated&measured\\
  parameter&&&\\
  &formula&value&value\\
  &&&\\\hline
  The velocity of zero-point&&&\\
  oscillations of&$\widehat{v_0}=c\alpha^3$&$116.5$&\\
  helium atom &&m/s&\\\hline\hline
  The density of atoms&&&\\
  in liquid &$n_3=\sqrt{\frac{M_3}{2m_e}}\frac{\alpha^2}{a_B^3}$&$1.88\cdot 10^{22}$&\\
  helium-3 &&$atom/cm^3$&\\\hline\hline
  The density&&&\\
  of liquid&$\gamma=M_3 n_3$&$93.7~$&$82.3~$\\
  helium-3, g/l&&&\\\hline\hline
  The dielectric&&&\\
  constant&$\frac{\varepsilon-1}{\varepsilon+2}=\frac{4\pi}{3}{\alpha^2}{\sqrt{\frac{M_3}{2m_e}}}$&1.035&\\
  of liquid helium-3&&&\\\hline\hline
   The boiling&&&\\
   temperature&$T_{boil}\simeq \frac{4}{3}\frac{\mathcal{E}_{x+y}}{k}$&$3.27$&$3.19$\\
  of helium-3,K&&&\\\hline\hline
   The sound velocity&&&\\
  in liquid&$c_{s}=2\widehat{v_0}$&233~&\\
  helium-3&&m/s&\\\hline\hline
  \end{tabular}
  \caption{The characteristic properties of liquid helium-3}
\label{DT3}
\end{table}

\bigskip

There is a radical difference between mechanisms of transition to the superfluid state for He-3 and He-4.
Superfluidity occurs if complete ordering exists in the atomic system.
For superfluidity of He-3 electromagnetic interaction should order not only zero-point vibrations of atoms, but also the magnetic moments of the nuclei.

It is important to note that  all characteristic dimensions of this task: the amplitude of the zero-point oscillations, the atomic radius, the distance between  atoms in liquid helium - all equal to the Bohr radius $a_B$ by the order of magnitude.  Due to this fact, we can estimate the oscillating magnetic field, which a fluctuating electronic shell creates  on "its" \ nucleus:
\begin{equation}
H_\Omega\approx\frac{e}{a_B^2}\frac{a_B \Omega_0}{c}\approx \frac{\mu_B}{A_3},
\end{equation}
where $\mu_B=\frac{e\hbar}{2m_ec}$ is the Bohr magneton, $A_3$ is the electric polarizability of helium-3 atom.

Because the value of magnetic moments for the nuclei He-3 is approximately equal to the nuclear Bohr magneton $\mu_{n_B}=\frac{e\hbar}{2m_pc}$, the ordering in their system must occur below the critical temperature
\begin{equation}
T_c = \frac{\mu_{n_B}H_\Omega}{k}\approx  10^{-3} K.
\end{equation}
This finding is in agreement with the measurement data.
The fact that the nuclear moments can be arranged in parallel or antiparallel to each other is consistent with the presence of the respective phases of superfluid helium-3.\\

Concluding this approach permits to explain the mechanism of superfluidity in liquid helium.

In this way, the  quantitative estimations of main parameters of the liquid helium and its transition to the superfluid state were  obtained.

 It was established that the phenomenon of superfluidity as well as the phenomenon of superconductivity
is based on the physical mechanism of the ordering of  zero-point oscillations.
\newpage

\chapter{Conclusion}
Until now it has been commonly thought that the existence of the isotope effect in superconductors leaves only one way for explanation  of the superconductivity phenomenon - the way based on the phonon mechanism.

Over fifty years of theory development based on the phonon mechanism, has not lead to success. All attempts to explain why some superconductors have certain critical temperatures (and critical magnetic fields) have failed.

This problem was further exacerbated with the discovery of high temperature superconductors. How can we move forward in HTSC understanding, if we cannot understand the mechanism that determines the critical temperature elementary superconductors?

 In recent decades, experimenters have shown that  isotopic substitution in metals leads to a change in the parameters of their crystal lattice and thereby affect the Fermi energy of the metal. As results, the  superconductivity can be based on a nonphonon mechanism.

The theory proposed in this paper suggests that the specificity of the association mechanism of electrons pairing is not essential.
It is merely important that such a mechanism was operational over the whole considered range of temperatures. The nature of the mechanism forming the electron pairs  does not matter, because although the work of this mechanism  is necessary it is still not a sufficient condition for the superconducting condensate's existence.
This is caused by the fact that after the electron pairing, they still remain as non-identical particles and  cannot form the condensate, because the individual pairs differ from each other as they commit uncorrelated zero-point oscillations.
 Only after an ordering of these zero-point oscillations, an energetically favorable lowering of the energy can be reached and a condensate at the level of minimum energy can then be formed.
Due to this reason the  ordering of  zero-point oscillations must be considered as the cause of the occurrence of superconductivity.

Therefore, the density of superconducting carriers and the critical temperature of a superconductor are determined by the Fermi energy of the metal, The critical magnetic field of a superconductor is given by the mechanism of destruction of the coherence of zero-point oscillations.

In conclusion, the consideration of zero-point oscillations allows us  to construct the theory of superconductivity, which is characterized by the ability to give estimations for the critical parameters of elementary superconductors. These results  are in satisfactory agreement with measured data.

This approach permit to explain the mechanism of superfluidity in liquid helium.
For electron shells  of atoms in S-states, the energy of interaction of zero-point oscillations  can be considered as a manifestation of  Van-der-Waals forces.
In this way the apposite  quantitative estimations of temperatures of the helium liquefaction  and its transition to the superfluid state was  obtained.

Thus it is established that both related phenomena, superconductivity and  superfluidity, are based on the same physical mechanism - they both are consequences of the ordering of  zero-point oscillations.


\end{document}